\documentclass[nofootinbib,twocolumn,aps,pre,superscriptaddress,citeautoscript,floatfix]{revtex4-1}
\usepackage{mathrsfs}
\usepackage{graphicx}

\usepackage{subcaption}

\usepackage{amsmath}
\usepackage{mathtools}
\usepackage{epstopdf}
\usepackage{braket}
\newcommand{\D}{{\rm d}}

\begin{document}

\title{Surface Pressure of Charged Colloids at the Air/Water Interface}
\author{Aviv Karnieli}
\affiliation{Raymond and Beverly Sackler School of Physics and Astronomy, Tel Aviv University, Ramat Aviv 69978, Tel Aviv, Israel}
\author{Tomer Markovich}
\affiliation{Raymond and Beverly Sackler School of Physics and Astronomy, Tel Aviv University, Ramat Aviv 69978, Tel Aviv, Israel}
\affiliation{DAMTP, Centre for Mathematical Sciences, University of Cambridge\\ Cambridge CB3 0WA, United Kingdom}
\author{David Andelman}\thanks{Corresponding author: andelman@post.tau.ac.il}
\affiliation{Raymond and Beverly Sackler School of Physics and Astronomy, Tel Aviv University, Ramat Aviv 69978, Tel Aviv, Israel}


\begin{abstract}
Charged colloidal monolayers at the interface between water and air (or oil) are used in a large number of chemical, physical and biological applications. Although a considerable experimental and theoretical effort has been devoted in the past few decades to investigate such monolayers, some of their fundamental properties are not yet fully understood. In this paper, we model charged colloidal monolayers as a continuum layer of finite thickness, with separate charge distribution on the water and air sides. The electrostatic surface free-energy and surface pressure are calculated via the charging method and within the Debye-H{\"u}ckel approximation. We obtain the dependence of surface pressure on several system parameters: the monolayer thickness, its distinct dielectric permittivity, and the ionic strength of the aqueous subphase. The surface pressure scaling with the area per particle, ${a}$, is found to be between ${a}^{-2}$ in the close-packing limit, and ${a}^{-5/2}$ in the loose-packing limit. In general, it is found that the surface-pressure is strongly influenced by charges on the air-side of the colloids. However, when the larger charge resides on the water-side, a more subtle dependence on salt concentration emerges. This corrects a common assumption that the charges on the water-side can \textit{always} be neglected due to screening. Finally, using a single fit parameter, our theory is found to fit well the experimental data for strong to intermediate strength electrolytes. We postulate that an anomalous scaling of $a^{-3/2}$, recently observed in low ionic concentrations, cannot be accounted for within a linear theory, and its explanation requires a fully-nonlinear analysis.
\end{abstract}
 
\maketitle

\section*{Introduction}

Molecular monolayers at the water/air or water/oil interfaces have been investigated intensively for more than a century, starting with the pioneering works of Langmuir and Blodgett~\cite{Langmuir1918,Blodgett1935,Vogel2012,Gaines, Adamson, Fundamentals, Binks2006}. Not only do they provide an important manifestation of thermodynamics of two-dimensional systems, but they equally offer several interesting applications in nano-lithography, micro-patterning and optical coatings~\cite{Vogel2012, Szekeres2002, Krogman2005}.

Related systems are monolayers of \textit{colloidal} particles deposited at fluid/fluid interfaces. Much interest in the latter systems followed the seminal work of Pieranski~\cite{Pieranski1980}, who  found that sub-micron polystyrene spheres are trapped at the air/water interface and self-assemble into a triangular lattice due to electrostatic repulsive interactions. More recently, a wide range of studies, including crystallization and aggregation of colloidal particles, have been performed on such monolayers~\cite{Onoda1985, Hurd1985A, Garcia1997,Reincke2004}.

Another key property of colloidal monolayers is their surface pressure/area isotherm. Such quantitative knowledge allows a direct control of particle spreading and self-assembling at the interface.  The surface pressure can be used to fine tune the inter-particle spacing when the monolayer is deposited from an aqueous solution~\cite{Mbamala2003}. Furthermore, from measurements of inter-particle forces and the monolayer surface pressure, one can infer the magnitude of the effective colloidal charge~\cite{Aveyard2000, Petkov2014, Petkov2016, Bossa2016}, as this quantity is otherwise hard to measure.

Several approaches have been suggested for calculating the surface pressure~\cite{Levental2008,Biesheuvel2007,Andelman1970,Hachisu1970,Aveyard2000,Petkov2014}. Levental~\textit{et al.}~\cite{Levental2008} as well as Biesheuvel and Soestbergen~\cite{Biesheuvel2007} modeled a charged monolayer as a surface with continuous and uniform charge density, and calculated the \textit{electrostatic} contribution to the surface pressure. Using the nonlinear Poisson-Boltzmann (PB) theory leads to a nonlinear expression for the surface pressure, $\Pi$, expressed in terms of hyperbolic functions. This result, when treated within the linearized PB theory (the Debye-H\"uckel theory -- DH, valid for small zeta potentials) yields a surface pressure that scales with $a$, the average area per particle, as $\Pi\sim{a}^{-2}$. In the opposite limit of weak electrolytes, however, the scaling is found to be $\Pi\sim{a}^{-1}$ \cite{Levental2008}. Since the model is only valid for a uniform surface charge density, it is restricted to the close-packing limit of the colloids, where the monolayer surface charge can be considered as approximately uniform.

In the other limit of large inter-particle separation, Aveyard~\textit{et al.}~\cite{Aveyard2000} studied the surface pressure of a charged monolayer at the water/oil interface, and calculated $\Pi$ using a superposition of inter-particle forces. These forces can be explained as a consequence of trapped charges residing at the particle/oil surface (in contact with the oil phase), away from the oil/water interface. The charges induce opposite image charges inside the aqueous phase, as a way to satisfy the dielectric discontinuous jump at the water/oil interface. The monolayer in the dilute limit can be treated as a dipolar layer and yields a surface pressure that scales as $\Pi\sim{a}^{-5/2}$. We note that the same scaling law was found to be in agreement with their own experimental results~\cite{Aveyard2000}.

The model by Aveyard~\textit{et al.}~\cite{Aveyard2000} mentioned above does not take into account the bulk concentration of ions in the aqueous sub-phase, and is a reasonable approximation only in the high-ionic strength (hence screened) limit. Moreover, as the model does not consider explicitly the distinct value of the dielectric constant of the colloidal monolayer, the dependence of $\Pi$ on the Debye screening length and monolayer dielectric constant have not been calculated.

Recently, Petkov~\textit{et al.}~\cite{Petkov2014,Petkov2016} measured the monolayer surface pressure, $\Pi$, for charged silica particles deposited at the air/water interface.
They calculated $\Pi$ from the Maxwell stress tensor, employing the so-called {\it Bakker formula}~\cite{Bakker}. The electrostatic field is assumed to vanish in the aqueous phase and was calculated in the air phase by postulating some specific ionic profiles. The surface charge density of each colloid and its accompanying screening was evaluated within a cell that is superimposed on a square lattice. In the large inter-particle separation limit, it was found that the surface pressure scales as $\Pi\sim{a}^{-3/2}$.

This scaling was shown to be in good agreement with experimental data~\cite{Petkov2014,Petkov2016,Vermant2006} measured either in the absence of salt or for weak electrolytes (using two fit parameters). It contradicts, however, the scaling law found earlier in ref~\cite{Aveyard2000}. Albeit the agreement between the model and experiment, the model was not derived in a self-consistent fashion. In particular, the use of the Bakker formula cannot be justified, because it relies on the homogeneity of the surface charge distribution, and the ionic profiles were postulated \textit{a priori} to have a preset form. In addition, the ansatz used to express the screening resembles the form of a typical solution in the DH (strong electrolytes) theory, although the considered experimental regime (weak electrolytes) is clearly beyond the scope of this approximation.

Motivated by these different models that yield distinct scaling forms ($\Pi\sim{a}^{-\alpha}$, where $\alpha=1,2$~\cite{Levental2008}, $5/2$~\cite{Aveyard2000} or $3/2$~\cite{Petkov2014}), we present in this paper a different, more fundamental and self-consistent approach. The thermodynamic definition of the surface pressure is employed without the need to have any further assumptions other than using the linearized DH theory.
Our calculation shows that for small inter-particle separation, the collective monolayer surface charge behaves as a continuum density, and the surface pressure scales as $\Pi\sim{a}^{-2}$, in agreement with ref~\cite{Levental2008}. For large separation, the colloids show a dipole-like behavior, and the scaling becomes $\Pi\sim{a}^{-5/2}$, recovering the results of ref~\cite{Aveyard2000}. In addition to the agreement with the two limiting scenarios of colloidal packing of previous works~\cite{Aveyard2000,Levental2008}, our model provides a general dependence on the entire area per particle range, $\Pi=\Pi(a)$. The theory derived herein also agrees well with available experimental data \cite{Aveyard2000,Vermant2006} within the DH regime.

The present study addresses a generalized setup, where the colloidal monolayer has a finite thickness and an arbitrary value of the effective dielectric constant. By generalizing previous works \cite{Levental2008,Biesheuvel2007,Andelman1970,	Petkov2016,Bossa2016,Hachisu1970,Aveyard2000,Petkov2014}, we allow the colloidal charge distribution to be different on the water-side versus the air-side of the monolayer. The surface pressure is obtained for any average inter-colloidal distance, and found to depend differently on the monolayer permittivity in the two limits of inter-particle separation. Furthermore, the dependence on the salt concentration can become non-monotonic for specific values of the charge density at the water-side. This finding is in contrast with the commonly employed assumption that the pressure depends solely on the charge located on the air-side of the colloids, for which the dielectric constant is much smaller, and when there is no screening.


\section*{Surface Pressure of a Charged Interface}
We present a general framework for calculating the surface pressure of an arbitrarily charged interface coupled to a bulk ionic solution. The definition of the surface tension, $\gamma$, is
\begin{equation}
\label{eq:Surface tension definition}
\gamma= \left( \frac{\partial F}{\partial A}\right)_{T,V} \, ,
\end{equation}
where $F$ is the free energy of the system (comprising an interface coupled with a bulk), and $A$ is the overall surface area.

The surface pressure is related to the change in surface tension.
For a charged surface, we can compare the surface tension with and without the charges
\begin{equation}
\label{eq:Surface Pressure definition}
\Pi= \gamma_0-\gamma\equiv -\Delta \gamma_{\mathrm{el}} \, ,
\end{equation}
where $\gamma_0$ and $\gamma$ denote the surface tension in the absence and presence of surface charges, respectively. The electrostatic contribution to the surface pressure, $\Pi$, is expressed in terms of $\Delta f_{\rm el}$, the change in electrostatic surface free-energy,
\begin{equation}
\label{eq:Surface Tension Non-Uniform}
\begin{multlined}
\Pi= -\left(\frac{\partial}{\partial A}\int_A \Delta f_{\mathrm{el}}\,\D^2r \right)_{T,V} \, .
\end{multlined}
\end{equation}
The surface free-energy, $\Delta f_{\rm el}$, is defined as the amount of work (per unit area) needed to \textit{construct} the surface. We introduce now the spatially averaged surface free-energy as
\begin{equation}
\label{eq:Surface Free Energy Non-Uniform}
\braket{\Delta f_{\mathrm{el}}}=\frac{1}{A}\int_{A} \Delta f_{\mathrm{el}}\,\D^2 r \,  ,
\end{equation}
and via eq~\ref{eq:Surface Tension Non-Uniform} write the surface pressure, in terms of the mean area \textit{per colloid}, $a\equiv A/N$, where $N$ is the number of colloidal particles on the surface,
\begin{equation}
\label{eq:Surface Pressure Non-Uniform}
{\Pi= -\braket{\Delta f_{\mathrm{el}}}-a\left(\frac{\partial \braket{\Delta f_{\mathrm{el}}}}{\partial a}\right)_{T,V}\, .}
\end{equation}

It is important to consider how the surface area is controlled in experiments. For a uniform surface charge density ({for which $\braket{\Delta f_{\mathrm{el}}}=\Delta f_{\mathrm{el}}$}), two fundamentally different situations can occur, and are known as the \textit{Gibbs monolayer} and the \textit{Langmuir monolayer}~\cite{Fundamentals}. For Gibbs monolayers, the particles are soluble in the aqueous sub-phase. The monolayer is an open system exchanging particles with the bulk, such that the chemical potential remains fixed. For a charged monolayer, it means that when the monolayer expands or contracts, its surface charge density remains constant, because the underlying physical properties that determine the surface coverage, such as the adsorption energy per unit area, approximately remain fixed~{\cite{Markovich2014,Markovich2015,Biesheuvel2007}}. As a result, $\braket{\Delta f_{\mathrm{el}}}$ is independent of the surface area and,
$\Pi=-\braket{\Delta f_{\mathrm{el}}}$, by virtue of eq~\ref{eq:Surface Pressure Non-Uniform}.

On the other hand, for Langmuir monolayers, the particles at the surface are completely insoluble in the water sub-phase, and the monolayer is a closed system with a fixed number of particles. The total monolayer charge, $Q=\int \sigma \,\D^2 r$,  remains fixed, meaning that $\sigma\sim a^{-2},$ even when the monolayer undergoes expansion or compression. For a uniformly charged surface and within the linear DH theory, the surface free-energy satisfies $\braket{\Delta f_{\mathrm{el}}}\sim a^{-2}$, and from eq~\ref{eq:Surface Pressure Non-Uniform}, $\Pi=\braket{\Delta f_{\mathrm{el}}}$.

Although these two simple cases may seem similar at first glance, the Gibbs and Langmuir monolayers yield an opposite relation between $\Pi$ and $\braket{\Delta f_{\mathrm{el}}}$, as shown above. Clearly, these two extreme cases of {\it uniform} surface charge density are an over-simplification, and for non-uniform surface charge densities, the relation between $\Pi$ and $\braket{\Delta f_{\mathrm{el}}}$ becomes more intricate.

In the present study, we only consider the case of insoluble Langmuir monolayers, where the total charge $Q$ (and particle number $N$) at the interface is constant but the charge density (per unit area) $\sigma$ can vary.
The charged surface is coupled to an electrolyte solution, and $\Delta f_{\mathrm{el}}$ is calculated using the Poisson-Boltzmann (PB) theory~\cite{Andelman1995, Markovich2017}.
The water and air phases are treated as two continuum media with dielectric constants, $\varepsilon_{\rm w}$ and $\varepsilon_{\rm a}$, respectively. The mobile ions in the aqueous solution are taken to be point-like, yielding the well-known PB equation for a monovalent 1:1 electrolyte
\begin{equation}
\label{eq:PB_eq}
\nabla^{2} \psi=\frac{2en_\mathrm{b}}{\varepsilon_0\varepsilon_{\rm w}}\sinh{\left(\frac{e\psi}{k_\mathrm{B} T}\right)} \, ,
\end{equation}
where $\psi$ is the electrostatic potential, $e$ the elementary charge, $\varepsilon_0$ the vacuum  permittivity, $n_\mathrm{b}$ the bulk concentration of the electrolyte and $k_\mathrm{B}T$ the thermal energy.

For an interface with surface charge $\sigma$ separating two media, the electrostatic boundary condition is
\begin{equation}
\label{eq:DH_BC}
\hat{\mathbf{n}}\cdot\left[\varepsilon_\mathrm{r}^{-}\nabla\psi^{-} - \varepsilon_\mathrm{r}^{+}\nabla\psi^{+}\right]=\frac{\sigma}{\varepsilon_0} \, ,
\end{equation}
where $\hat{\mathbf{n}}$ denotes the unit vector normal to the surface. The $\pm$ superscripts on the potential and relative permittivity $\epsilon_{\mathrm{r}}$ of the media (\textit{e.g.}, $\varepsilon_{\rm a}$, $\varepsilon_{\rm w}$, etc.) denote the external ($+$) and internal ($-$) regions with respect to the surface, and the direction of $\hat{\mathbf{n}}$ is chosen to point from inside toward the outside.

From the well known \textit{charging method}~\cite{Andelman1995,Overbeek1948, Markovich2017}, the electrostatic free-energy due to the presence of an electric double layer can be evaluated as
\begin{equation}
\label{eq:Surface Free Energy DLVO}
\Delta f_{\mathrm{el}}= \int_{0}^{\sigma} \psi_\mathrm{s} (\sigma')\,\D\sigma' \, .
\end{equation}
Eqs~\ref{eq:Surface Free Energy Non-Uniform} through \ref{eq:Surface Free Energy DLVO} are sufficient to obtain the surface pressure in the most general case\footnote{Although the above charging method takes into account the entropy of the mobile ions in the solution, it does not include the entropy of the surface charges. In our model, those charges originate from the charge distribution on large colloidal particles, forming a monolayer at the air/water interface. However, since the colloids are considered as macro-particles, this entropy contribution can be ignored.}. However, for simplicity, the Debye-H{\"u}ckel (DH) linearization scheme can be employed~\cite{Andelman1995} for eq~\ref{eq:PB_eq},
\begin{equation}
\label{eq:DH_eq}
\nabla^{2} \psi=\kappa_\mathrm{D} ^{2}\psi \,  ,
\end{equation}
for cases when $\psi \ll k_\mathrm{B} T /e$.
In the above equation, the inverse Debye screening length is $\kappa_\mathrm{D}=\sqrt{8\pi l_\mathrm{B}n_\mathrm{b}}$, and the Bjeruum length $l_{\mathrm{B}}=e^2/(4\pi\varepsilon_0\varepsilon_{\rm w} k_\mathrm{B} T)$ is about $0.7$\,nm in water ($\varepsilon_{\rm w}\simeq78$) at room temperature.

In the DH regime, $\psi_\mathrm{s}$ and $\sigma$ are related linearly~\cite{Andelman1995,Overbeek1948}, and eq~\ref{eq:Surface Free Energy DLVO} becomes
\begin{equation}
\label{eq:Surface Free Energy DLVO linear}
\Delta f_{\mathrm{el}}= \frac{1}{2} \psi_\mathrm{s} \sigma \, .
\end{equation}
This formula can be generalized for two charged surfaces, $S_i,~i=1,2$, each with surface potential, $\psi_{\mathrm{s},i}$, and surface charge, $\sigma_i$,
\begin{equation}
\label{eq:Surface free energy density linear}
\Delta f_{\mathrm{el}} =\frac{1}{2}\sum_{i=1,2}  \psi_{\mathrm{s},i} \sigma_i \, .
\end{equation}
Eq~\ref{eq:Surface free energy density linear} is obtained by the linearization of the expression presented in ref~\cite{BenYaakov2013}, in the context of two interacting charged surfaces.
In the following section, the above equation will be used to calculate the free energy for Langmuir monolayers, which are modeled as two charged interfaces separated by a dielectric layer (Figure~\ref{fig:Fig1}b).

\section*{Modeling of the Colloidal Monolayer}

\begin{figure}
	\centering
	\includegraphics[scale=0.25,draft=false]{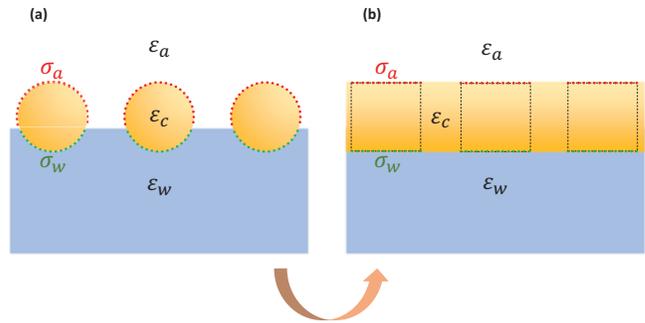}
	\caption{\textsf{(color online)~ } $\sigma_i$ and $\varepsilon_i$ refer to the surface charge on interface $i$ and permittivity in medium $i$, respectively. (a) Cross-section of a colloidal monolayer at the air/water interface. Spherical colloids are partly immersed in the aqueous phase, with different charge distribution on their air- and water-facing sides. (b) A simplified model, where the colloids are now cubes with surface charges residing only on the air- and water-facing facets. The monolayer region has a uniform effective dielectric constant, $\varepsilon_{\mathrm{c}}$.}
	\label{fig:Fig1}
\end{figure}

{We first consider a model for a monolayer of charged colloidal particles at the air/water interface as presented in Figure~\ref{fig:Fig1}a. Micron-sized colloids are modeled as dielectric spheres (of permittivity $\varepsilon_{\mathrm{c}}$) partially submerged in the aqueous phase. Different charge distributions are present on the colloids water- and air- facing surfaces, which, together with the ions in the electrolyte solution, give rise to the electrostatic interactions. Note that the colloids finite-size dictates an excluded volume for the compression. This complex geometry, however, hinders the simplicity of our surface free-energy method.}

{The aforementioned setup can be much simplified, when the partially immersed spherical colloids are modeled as dielectric cubes of same dimensions, and the surface charges are now positioned on the water- and air- facing facets of the cubes (Figure~\ref{fig:Fig1}b). Such approximation recovers the main physical features. The redistribution of charge merely introduces geometric corrections (as was similarly approached in refs~\cite{Petkov2014,Petkov2016,Bossa2016,Frydel2007}).}

{Moreover, we replace the dielectric constant as seen by the colloids with an effective permittivity $\varepsilon_{\mathrm{eff}}$, smeared over the monolayer region. This effective permittivity is comparbale to the permittivity of the colloids, and for simplicity is taken as $\varepsilon_{\mathrm{eff}}=\varepsilon_{\mathrm{c}}$. For close-packing of colloids, this represents a reasonable approximation. For loosely-packed monolayers, we find that the interaction is dictated by dipole-dipole forces mediated primarily through the \textit{air} phase. The role played by $\varepsilon_{\mathrm{eff}}$ is to merely determine the dipole strength, which is a local feature of the colloidal particle itself, consistent with our approximation $\varepsilon_{\mathrm{eff}}=\varepsilon_{\mathrm{c}}$.}

\begin{figure*}
	\qquad
	\begin{subfigure}{0.35\textwidth}
		\subcaption{}
		\includegraphics[width=\textwidth, draft=false,trim={1.5cm, 2cm, 1.5cm, 2cm}, clip]{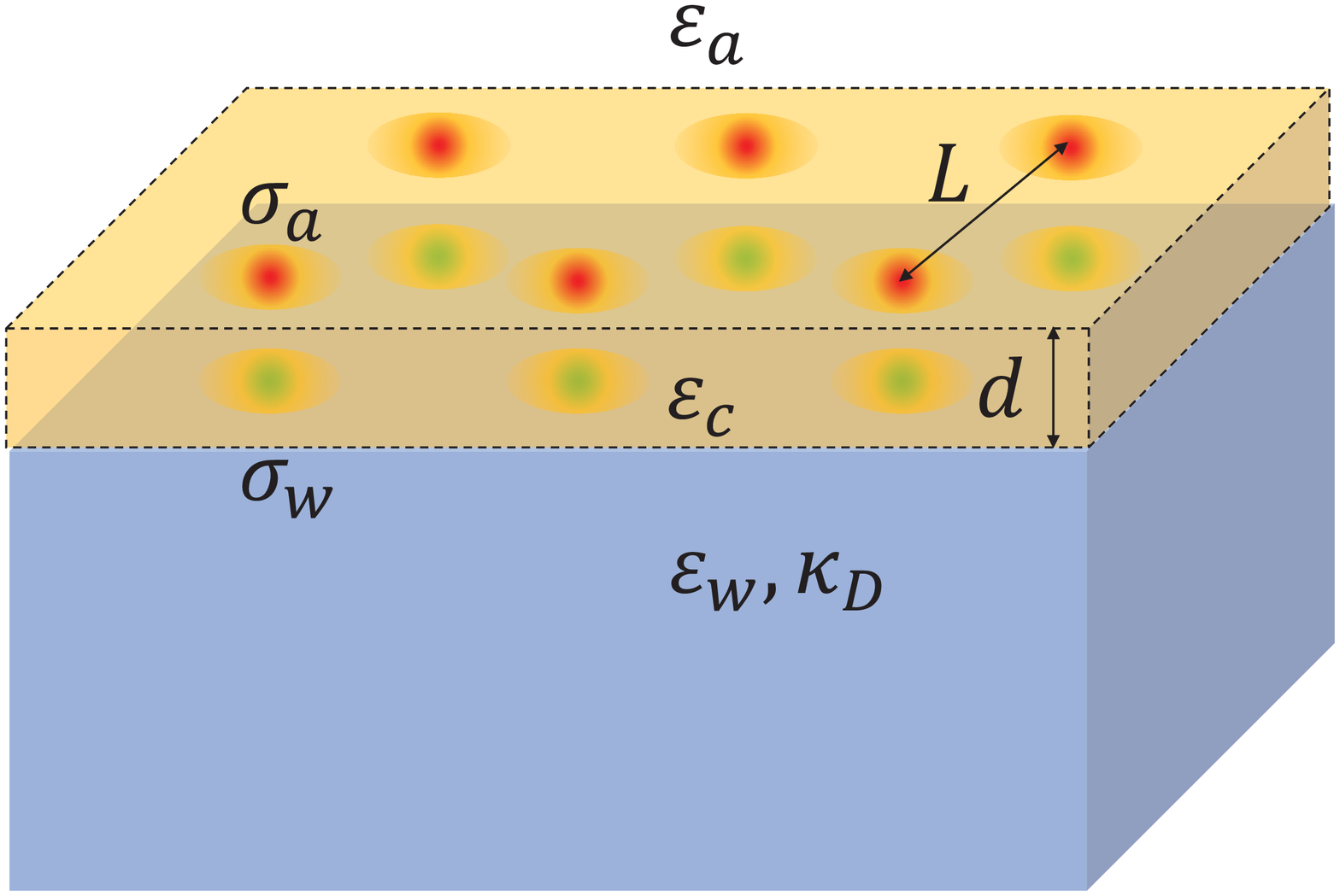}
		\label{fig:Fig2a}
	\end{subfigure}\qquad \qquad \qquad \qquad \quad
	\begin{subfigure}{0.35\textwidth}
		\centering
		\subcaption{}
		\includegraphics[width=\textwidth, draft=false, trim={2.5cm, 5cm, 2cm, 4.5cm}, clip]{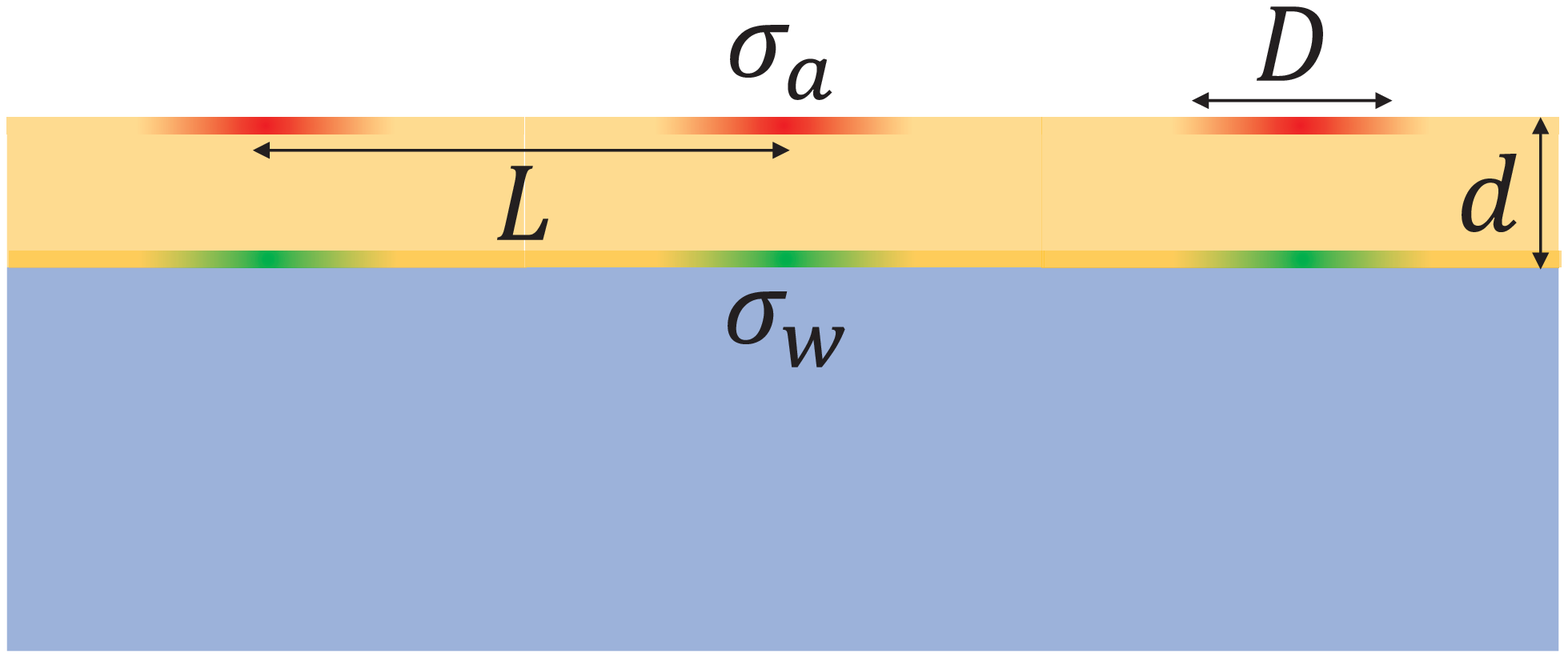} 
		\label{fig:Fig2b}
	\end{subfigure}
	\caption{\textsf{(color online) Schematic drawing of the colloidal monolayer in between two interfaces: a top one, at $z=0$, in contact with air (dielectric constant $\varepsilon_{\rm a}$), and a bottom one, at $z=-d$, in contact with an aqueous solution having dielectric constant, $\varepsilon_{\rm w}$, and Debye screening length, $\kappa_{\rm D}^{-1}$. (a) The layer of finite thickness, $d$,  is modeled as a dielectric layer of dielectric constant $\varepsilon_\mathrm{c}$,  spread between the water and air phases. The colloidal charges accumulate on the water-side and air-side, and form a square lattice with lattice constant $L$. The two corresponding surface charge densities are denoted, respectively, as $\sigma_{\rm a}$ and $\sigma_{\rm w}$. (b) Cross-section of the monolayer. The colloid charge density is spread over the colloid diameter, $D$, and the distance between colloid centers is $L$. }}
	\label{fig:Fig2}
\end{figure*}

The charged colloidal monolayer is schematically depicted in Figure~\ref{fig:Fig2}. It is located at the interface between a dielectric medium of constant $\varepsilon_{\rm a}$ (non-aqueous medium such as air or oil) on its top side, and a monovalent 1:1 electrolyte solution, of dielectric constant $\varepsilon_{\mathrm{w}}$, on its bottom side. The colloidal monolayer medium is considered continuous with finite thickness, $d$ (we later set it equal to the colloid diameter, thus ignoring immersion in the aqueous phase due to wetting), and a dielectric constant, $\varepsilon_\mathrm{c}$. The charges residing on each side of a single colloid are modelled as a patch of surface charge (see also Figure~\ref{fig:Fig1}). The charge distribution can take different values on the air-facing ($z=0$) and water-facing ($z=-d$) sides.

The monolayer itself is constructed by repeating the pairs of surface charge patches (each representing an \textit{individual} colloid) on a square lattice with lattice parameter $L$, as is seen in Figure~\ref{fig:Fig2}. Here, $L$ is the average distance between the colloids, which takes into account the colloids excluded volume. 
Note that the patches can have an arbitrary shape and charge distribution with typical length scale, $D<L$, which serves as an effective colloid diameter. For an arrangement on a square lattice, the total surface charge densities on the monolayer air-side and water-side, $\sigma_{\rm a}(x,y)$ and $\sigma_{\rm w}(x,y)$, become periodic functions in $x\to x+L$ and $y\to y+L$.

In order to calculate the surface pressure, we need to evaluate first the electrostatic potential. The potential in the air phase, $\psi^{(\mathrm{a})}$,
as well as inside the colloidal monolayer region, $\psi^{(\mathrm{c})}$, satisfies the Laplace equation,
\begin{equation}
\label{eq:Laplace Equation1}
\nabla^{2} \psi^{(\mathrm{a})}=0\, , ~~~
\nabla^{2} \psi^{(\mathrm{c})}=0 \, ,
\end{equation}
while the potential in the aqueous phase, $\psi^{(\mathrm{w})}$, satisfies the DH equation, eq~\ref{eq:DH_eq}
\begin{equation}
\label{eq:DH Equation}
\nabla^{2} \psi^{(\mathrm{w})}=\kappa_\mathrm{D} ^{2}\psi^{(\mathrm{w})} \, .
\end{equation}
and depends on the solution ionic strength via the Debye screening length, $\kappa_{\rm D}^{-1}$.

The boundary conditions at the $z=0$ and $z=-d$ planes are obtained from eq~\ref{eq:DH_BC},
\begin{equation}
\nonumber
\label{eq:Vector_BC1_eq}
\varepsilon_\mathrm{c}\frac{\partial\psi^{(\mathrm{c})}}{\partial z}\Big{|}_{z=0^-} -\, \varepsilon_{\rm a}\frac{\partial\psi^{(\mathrm{a})}}{\partial z}\Big{|}_{z=0^+}=\frac{1}{\varepsilon_0}\sigma_{\rm a}(x,y) \, ,
\end{equation}
\begin{equation}
\label{eq:Vector_BC2_eq}
\varepsilon_{\rm w}\frac{\partial\psi^{(\mathrm{w})}}{\partial z}\Big{|}_{z=-d^-} -\, \varepsilon_\mathrm{c}\frac{\partial\psi^{(\mathrm{c})}}{\partial z}\Big{|}_{z=-d^+}=\frac{1}{\varepsilon_0}\sigma_{\rm w}(x,y) \, ,
\end{equation}
and at $z\to \pm \infty$ we demand that the electrostatic field vanishes, $\lim_{z\rightarrow\pm\infty}|\nabla\psi|=0$.

\section*{Results}

{The surface free-energy and pressure can be calculated (both numerically and analytically) by using the model periodicity and using the Fourier transform on the surface charges and the electric potential. The two periodic charge densities, $\sigma_{\rm a}(x,y)$ and $\sigma_{\rm w}(x,y)$, are expressed by their Fourier series
	\begin{eqnarray}
	\label{eq:Periodic Charge Fourier}
	\sigma_{\rm a}(\mathbf{r})&=&\frac{1}{L^2}\sum_{n,m=-\infty}^{\infty}\tilde{s}_{\rm a}(\mathbf{k})
	\mathrm{e}^{i\mathbf{k}\cdot\mathbf{r}} \nonumber \, , \\
	\sigma_{\rm w}(\mathbf{r})&=&\frac{1}{L^2}\sum_{n,m=-\infty}^{\infty}\tilde{s}_{\rm w}(\mathbf{k})
	\mathrm{e}^{i\mathbf{k}\cdot\mathbf{r}} \, ,
	\end{eqnarray}
where $\mathbf{r}=(x,y)$ is the in-plane vector. For a square lattice, the integer numbers $\{n,\,m\}=0,\pm 1,\pm 2\dots,$ parameterize the discrete $\mathbf{k}$-vector of the reciprocal space with $\mathbf{k}=2\pi /(L \sqrt{n^2+m^2})$, and $\tilde{s_i}(\mathbf{k})$, $i={\rm a,w}$, is the $\mathbf{k}$-component of the Fourier transform of a {\it single} colloid charge distribution. In a similar manner to eq~\ref{eq:Periodic Charge Fourier}, the surface potentials evaluated at the top and bottom surfaces, $\psi_{\mathrm{s}}^{(\mathrm{a})}$ and $\psi_{\mathrm{s}}^{(\mathrm{w})}$, can also be written in terms of their Fourier components:
\begin{equation}
\label{eq:Periodic Potential Fourier}
\psi_{\mathrm{s}}^{(i)}(\mathbf{r})=\sum_{n,m=
	-\infty}^{\infty}\tilde{\psi}_{\mathrm{s}}^{(i)}(\mathbf{k})\mathrm{e}^{i\mathbf{k}\cdot\mathbf{r}} \, ,
\end{equation}
with $i={\rm a,w}$, and $\tilde{\psi}_{\mathrm{s}}^{(i)}(\mathbf{k})$ denotes the Fourier coefficients of $\psi_{\mathrm{s}}^{(i)}(\mathbf{r})$.}

{A linear relation between the surface potential and surface charge density emerges from the boundary conditions as our model is linear. The generalized linear response is written in Fourier space as the product}
\begin{equation}
\label{eq:Vector Linear Relation between surface potential and surface charge}
{\tilde{\Psi}_\mathrm{s}(\mathbf{k})=\frac{1}{L^2}\mathrm{C^{-1}}(\mathbf{k})\tilde{\Sigma} (\mathbf{k}) \, ,}
\end{equation}
{where in a compact notation, $\tilde{\Psi}_\mathrm{s}(\mathbf{k})=
(\tilde{\psi}_{\mathrm{s}}^{(\mathrm{a})}(\mathbf{k}),\tilde{\psi}_{\mathrm{s}}^{(\mathrm{w})}(\mathbf{k}))$ and $\tilde{\Sigma}(\mathbf{k})=(\tilde{s}_{\rm a}(\mathbf{k}),\tilde{s}_{\rm w}(\mathbf{k}))$ are vectors, and $\mathrm{C}(\mathbf{k})$ is a $2\times 2$  matrix. The diagonal components of the $\mathrm{C}(\mathbf{k})$ matrix are the differential capacitances (per unit area) of the `a' and `w' surfaces, while the off-diagonal ones represent cross-capacitance terms between the two surfaces. One can then write $\braket{\Delta f_{\rm el}}$ as a function of the inverse capacitance and the surface charge, by using eq~\ref{eq:Surface free energy density linear} and Parseval's theorem,
\begin{equation}
\label{eq:Vector Surface Pressure Explicit 1}
\begin{multlined}
\braket{\Delta f_{\rm el}}=\frac{1}{2L^4}\sum_{\mathbf{k}}\tilde{\Sigma}\mathrm{C^{-1}}\tilde{\Sigma}\, .
\end{multlined}
\end{equation}}
{Since $\braket{\Delta f_{\rm el}}$ depends explicitly on the area per colloid $a=L^2$, eq~\ref{eq:Surface Pressure Non-Uniform} can be employed to calculate the surface pressure (see Appendix A for details).}

{The quantities appearing in eq~\ref{eq:Vector Surface Pressure Explicit 1} are all obtained from the explicit solution of the boundary value problem defined in Section III for the electrostatic potential (see Appendix A for details). In order to obtain the potential, the boundary conditions (\textit{i.e.}, the surface charge distributions) must be specified. Here, we assume that this charge distribution is Gaussian, and obtains separate values on the air (`a') and water (`w') sides of the monolayer}
\begin{equation}
\label{eq:Gauss}
{s(r)= \frac{2Q}{\pi D^2}\exp{\left(-\frac{2r^2}{D^2}\right)} \, .}
\end{equation}
{where $r=|\boldsymbol{r}|$, and $Q=Q_{\rm a}$ or $Q_{\rm w}$ is the charge on the `a' or `w' sides, respectively. The limiting values for the surface pressure, however, still remain independent of this specific choice of profile.}
\subsection*{Close-packing colloid limit, {$L \to  D$}}
{In the close-packing limit, the inter-particle spacing $L$ approaches the colloid effective diameter $D$. In this case, the limiting value of the surface pressure $\Pi$ can be derived analytically (see Appendix A for details)}

\begin{equation}
\label{eq:Pi_cp_text}
{\Pi\simeq \left[\frac{Q_{\rm a}^2d}{2\varepsilon_\mathrm{c}\varepsilon_0}
+\frac{(Q_{\rm a}+Q_{\rm w})^2}{2\varepsilon_0\varepsilon_{\rm w}\kappa_\mathrm{D}}\right]\frac{1}{{a}^2},}
\end{equation}
where $a\equiv L^2$ is the area per colloid. Therefore, in the close-packing limit $\Pi$ scales as $a^{-2}$, and is consistent with the continuum limit of the monolayer surface charge. As the colloids approach each other, their double-layers overlap and resemble the response to a uniform surface charge density at the air/water interface. Note that eq~\ref{eq:Pi_cp_text} is independent of the specific surface charge distribution, $s(\mathbf{r})$ of each colloid.

\subsection*{Dilute colloid limit, {$L \gg D$}}
{The opposite regime of inter-particle separation is the dilute limit, where $L\gg D$. A closed-form analytic expression for $\Pi$ can be derived by using the Euler-Maclaurin formula (see Appendix A for details). The result obtained suggests that the surface pressure originates from dipolar interactions between the colloids:}
\begin{equation}
\label{eq:Pi_dipole}
{\Pi=\frac{\pi}{12}\frac{p^2_{\mathrm{eff}}}{\varepsilon_0\varepsilon_{\rm a}}
\frac{1}{{a}^{5/2}}\,+\,O\left(\frac{1}{{a}^{7/2}}\right) \, ,}
\end{equation}
where the effective dipole moment $p_{\mathrm{eff}}$ is written as the sum of two terms,
\begin{equation}
\label{eq:dipole}
p_{\mathrm{eff}}=p_1+p_2= \frac{2\varepsilon_{\rm a}}{\varepsilon_\mathrm{c}}Q_{\rm a}d
\,+\,\frac{2\varepsilon_{\rm a}}{\varepsilon_{\rm w}}\frac{Q_{\rm a}+Q_{\rm w}}{\kappa_\mathrm{D}} \, .
\end{equation}

In Appendix~B, we calculate $\Pi$ directly from an effective model of dipole-dipole interactions and arrive at an identical result. This approach suggests that the dilute limit, as in eqs~\ref{eq:Pi_dipole}-\ref{eq:dipole}, is independent of the specific functional form of the surface charge density. One might indeed expect this behavior as the inter-particle distance satisfies $D\ll L$, and the details of the colloidal charge distribution are washed out.

We note that there are two separate cases for the effective dipole moment, eq~\ref{eq:dipole}:

(i)~In the case of strong electrolytes, where $\kappa_\mathrm{D} d\gg\left(1+Q_{\rm w}/Q_{\rm a}\right)(\varepsilon_\mathrm{c}/\varepsilon_{\rm w})$, $p_{\mathrm{eff}}$ is approximated as,
\begin{equation}
\label{eq:dipole_Qa}
p_{\mathrm{eff}}\simeq p_1= \frac{2\varepsilon_{\rm a}}{\varepsilon_\mathrm{c}}Q_{\rm a}d \, ,
\end{equation}
and mainly depends on $p_1$, since the air-exposed charge induces an image charge in the aqueous phase, in a distance of $2(\varepsilon_{\rm a}/\varepsilon_{\rm c})d$.

(ii)~In the case of weak electrolytes, namely, $\kappa_\mathrm{D} d\ll\left(1+Q_{\rm w}/Q_{\rm a}\right)(\varepsilon_\mathrm{c}/\varepsilon_{\rm w})$, $p_{\mathrm{eff}}$ is well approximated by the second term, $p_2$,
\begin{equation}
\label{eq:dipole_Qw}
p_{\mathrm{eff}}\simeq p_2=\frac{2\varepsilon_{\rm a}}{\varepsilon_{\rm w}}(Q_{\rm a}+Q_{\rm w}) \kappa_\mathrm{D} ^{-1} \, .
\end{equation}
The relevant charge determining the dipole moment is the net charge on both sides of the particle, as a result of the electro-neutrality due to screening. The charge separation is then proportional to the screening length.

For a given inter-particle separation $L$, we express the dependence of $\Pi$ on the ionic strength using eqs~\ref{eq:Pi_dipole}-\ref{eq:dipole},
\begin{equation}
\label{eq:Pi_Salt}
\Pi=\left[1+\left(1+\frac{Q_{\rm w}}{Q_{\rm a}}\right)\frac{\varepsilon_\mathrm{c}}{\varepsilon_{\rm w}}\frac{1}{\kappa_\mathrm{D} d}\right]^2 \Pi _{\infty} \, ,
\end{equation}
where $\Pi_{\infty}=\Pi(\kappa_\mathrm{D} d \to \infty)$ is the surface pressure for a vanishing screening length. From eq~\ref{eq:Pi_Salt}, we deduce that for $Q_{\rm w}/Q_{\rm a}>0$, $\Pi$ is a monotonic function of the Debye screening length, and converges to $\Pi_{\infty}$ for very high electrolyte concentrations.

\begin{figure}
	\centering
	\includegraphics[scale=0.45,draft=false]{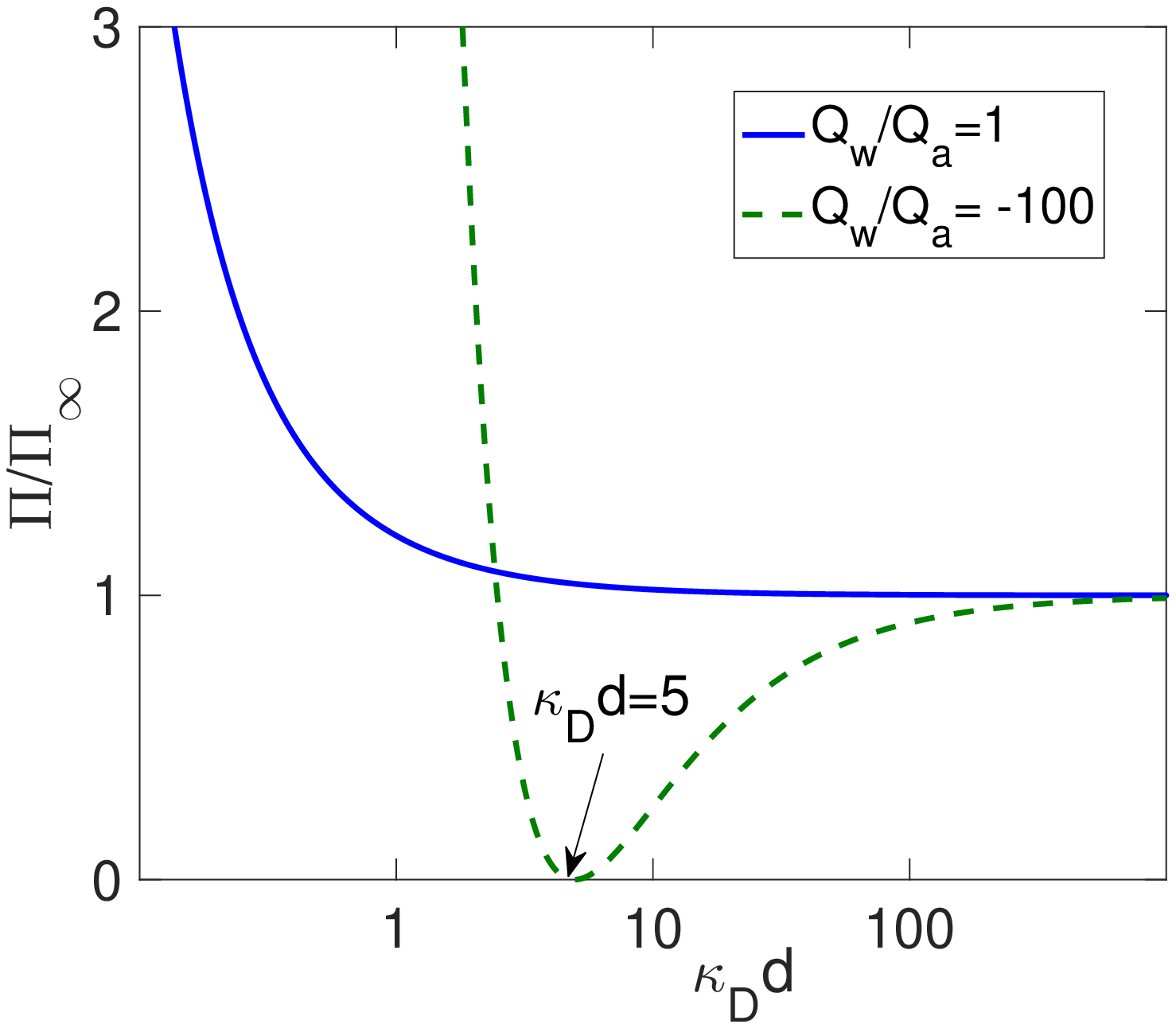}
	\caption{\textsf{(color online)~ Rescaled surface pressure, $\Pi/\Pi_{\infty}$, for the dilute limit ($\xi\ll1$), plotted as a function of the reduced screening parameter, $\kappa_\mathrm{D} d$. The rescaling factor, $\Pi_{\infty}=\Pi(\kappa_\mathrm{D} d\to \infty)$, is the pressure for strong electrolytes. Taking $\varepsilon_\mathrm{c}=4$ for silica particles and $\varepsilon_{\rm w}=80$, we compare the dependence on $\kappa_D d$ for two values of the charge ratio. For $Q_{\rm w}/Q_{\rm a}=1$ (blue solid line), the dependence is monotonic and does not vanish, while for negative and large ratios, $Q_{\rm w}/Q_{\rm a}=-100$ (dashed green line), the surface pressure is non-monotonic and even vanishes for a certain value of $\kappa_\mathrm{D} d$.}}
	\label{fig:Fig3}
\end{figure}

However, if the ratio $Q_{\rm w}/Q_{\rm a}$ becomes negative, $\Pi$ might even vanish for certain values of $\kappa_\mathrm{D}$, as can be seen in Figure~\ref{fig:Fig3} for a specific choice of $Q_{\rm w}/Q_{\rm a}=-100$. Moreover, $\Pi$ is non-monotonic with respect to the salt concentration. This presents a compelling evidence that surface charges on {\it both} sides of the colloid particle can play a role in determining the magnitude of $\Pi$. We shall further discuss this observation below.

\subsection*{{General inter-particle separations}}
The surface pressure, $\Pi$, can be calculated numerically for any intermediate value of inter-particle separation $L$. In this case it is most convenient to define a dimensionless parameter $\xi\equiv D/L$, where $0<\xi\leq 1$, with $\xi \to 1$ denoting the close-packing limit and $\xi \ll 1$ the dilute regime. We restrict ourselves to the more common case of strong electrolytes, $\kappa_D d\gg(1+Q_{\rm w}/Q_{\rm a})(\varepsilon_{\rm c}/\varepsilon_{\rm w})$, with $Q_{\rm w}=0$, $d=D$ and $\kappa_\mathrm{D} d=10$. We have chosen $Q_{\rm w}=0$ on the water side, without loss of generality, because it merely sets the strong electrolyte regime to $\kappa_D d \gg \varepsilon_{\rm c}/\varepsilon_{\rm w}$.

{The average electrostatic surface free-energy, $\braket{\Delta f_{\mathrm{el}}}$} is calculated by summing one hundred terms of the two series in eq~\ref{eq:Gxi}, where their explicit form is also given in Appendix~A. $\Pi(\xi)$ is then evaluated via eq~\ref{eq:Surface Pressure xi}.

\begin{figure}
	\includegraphics[scale=0.55,draft=false]{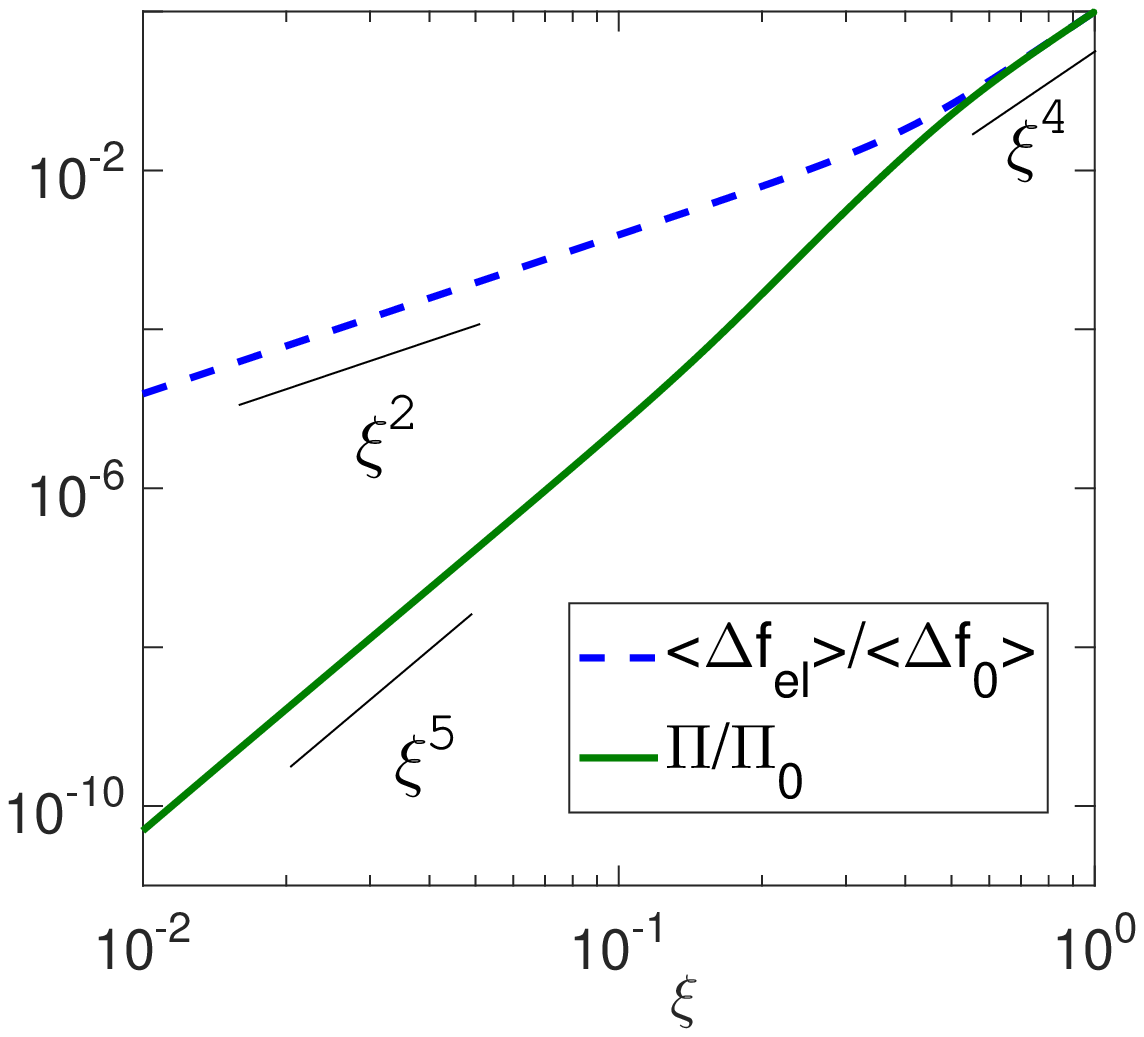}
	\caption{\textsf{(color online)~Rescaled surface pressure, $\Pi/\Pi_0$ (green solid line), and rescaled surface free-energy, $\braket{\Delta f_{\rm el}}/\braket{\Delta f_{0}}$ (blue dashed line), plotted on a log-log plot as a function of $\xi{=D/L}$. The rescaling is done with respect to the close-packing values at $\xi=1$. The free energy and pressure scale identically as $\sim\xi^4$ in the  $\xi \to 1$ limit, where the monolayer can be regarded as a having a uniform surface charge. In the dilute limit ($\xi \ll 1$), however, $\Pi\sim \xi^5$, and differs significantly from $\braket{\Delta f_{\mathrm{el}}}\sim \xi^2$. }}
	\label{fig:Fig4}
\end{figure}

The quantities $\braket{\Delta f_{\mathrm{el}}}$ and $\Pi$, rescaled by their maximal values at $\xi=1$ ($\Pi_0$ and $\braket{\Delta f_{0}}$, respectively), are shown in Figure~\ref{fig:Fig4} on a log-log plot. Clearly, both coincide in the close-packing limit ($\xi\lesssim 1$), where the continuum limit holds, {\it i.e.}, $\Pi\simeq \braket{\Delta f_{\mathrm{el}}}\sim\xi^4\sim{a}^{-2}$. However, in the dilute regime ($\xi\ll 1$), the surface free-energy and surface pressure differ considerably as $\braket{\Delta f_{\mathrm{el}}}\sim \xi^2 \sim{a}^{-1}$ and $\Pi\sim\xi^5 \sim{a}^{-5/2}$.

\begin{figure}
	\includegraphics[scale=0.55,draft=false]{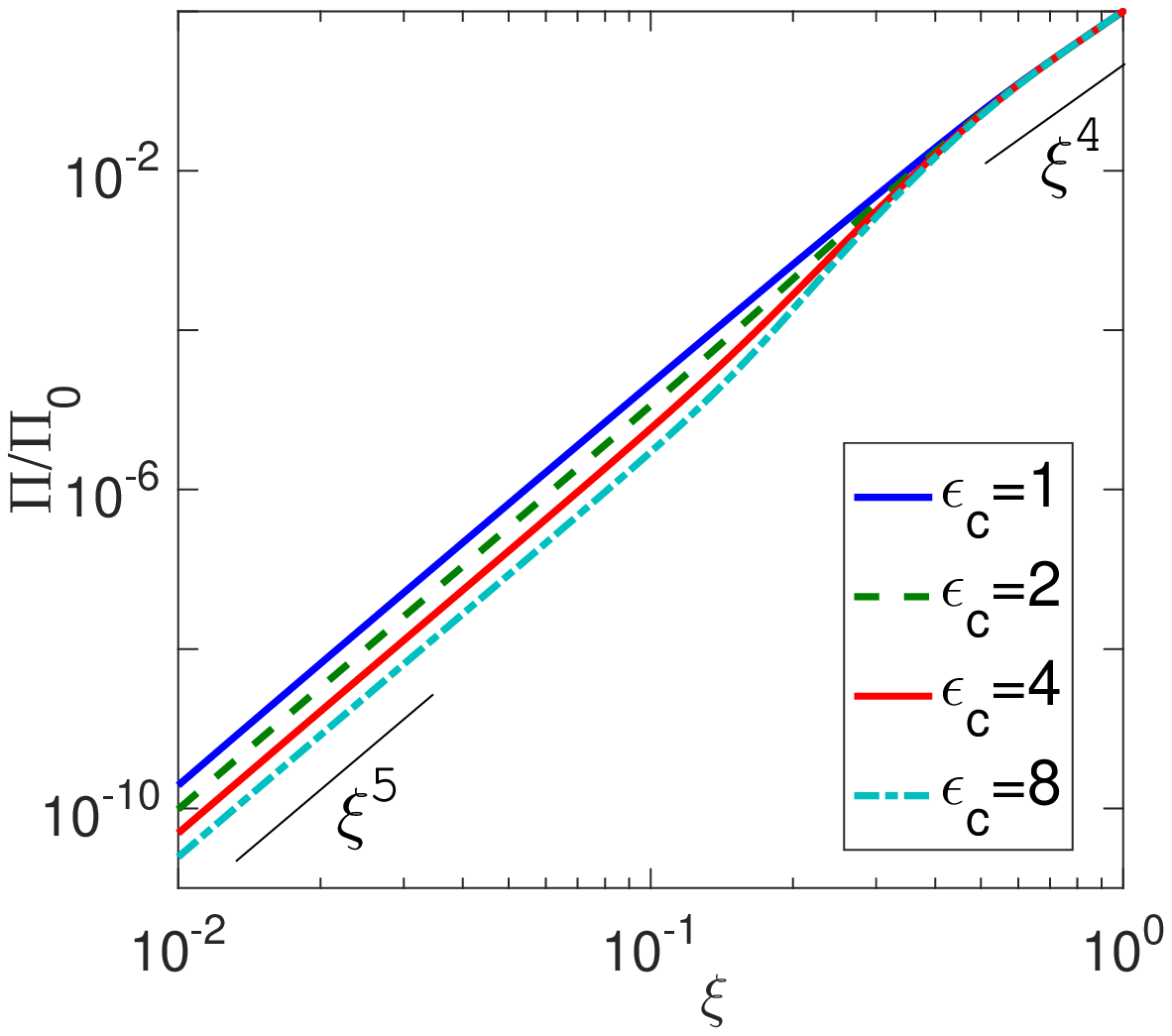}
	\caption{\textsf{(color online)~Log-log plot of the rescaled surface pressure, $\Pi/\Pi_0$ with $\Pi_0=\Pi(\xi=1)$, as a function of $\xi{=D/L}$, for different values of monolayer dielectric constant, $\varepsilon_\mathrm{c}=1, 2, 4,$ and 8. Two scaling regimes of $\xi^{5}$ and $\xi^4$ can be seen, as in figure~\ref{fig:Fig4}. The rescaled surface pressure, $\Pi/\Pi_0$, does not show any dependence on $\varepsilon_\mathrm{c}$ in the close-packing limit. In the dilute-packing limit, however, $\Pi/\Pi_0$ scales as $1/\varepsilon_\mathrm{c}$.}
		\label{fig:Fig5}}
\end{figure}

A plot of $\Pi$, rescaled by its maximal value, is given in Figure~\ref{fig:Fig5} for different values of the monolayer dielectric constant, $\varepsilon_\mathrm{c}=1,2,4,$ and $8$. The variation of the rescaled surface pressure with $\varepsilon_\mathrm{c}$ is quite substantial only in the dilute-packing limit, where it varies as $1/\varepsilon_\mathrm{c}$, as is implied by eq~\ref{eq:Pi_lp}. We note that $\Pi_0$ (the rescaling prefactor) is the close-packing value of $\Pi$, $\Pi_0=\Pi(\xi=1)$, and is by itself proportional to $1/\varepsilon_\mathrm{c}$, [see eq~\ref{eq:Pi_cp_text}].

\section*{Comparison to experiments}

\begin{figure}[h!]
	\captionsetup[subfigure]{labelformat=empty}
	\begin{subfigure}{0.45\textwidth}
		\subcaption{}
		\includegraphics[width=\textwidth,draft=false,trim={1cm, 0cm, 0cm, 0.5cm}, clip]{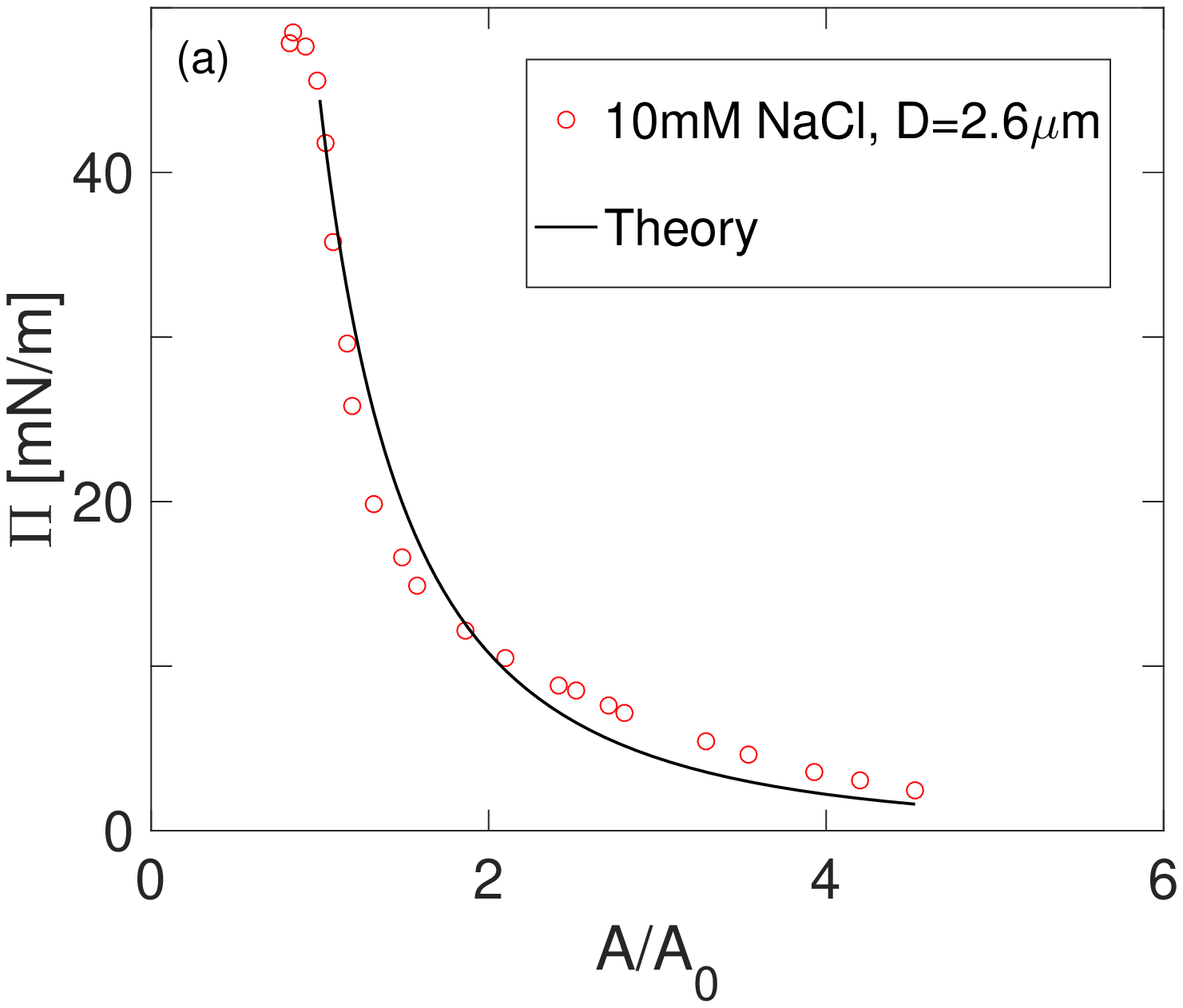}
		\label{fig:Fig6a}
	\end{subfigure}
	\begin{subfigure}{0.45\textwidth}
		\subcaption{}
		\includegraphics[width=\textwidth,draft=false,  trim={1cm, 0cm, 0cm, 0.5cm}, clip]{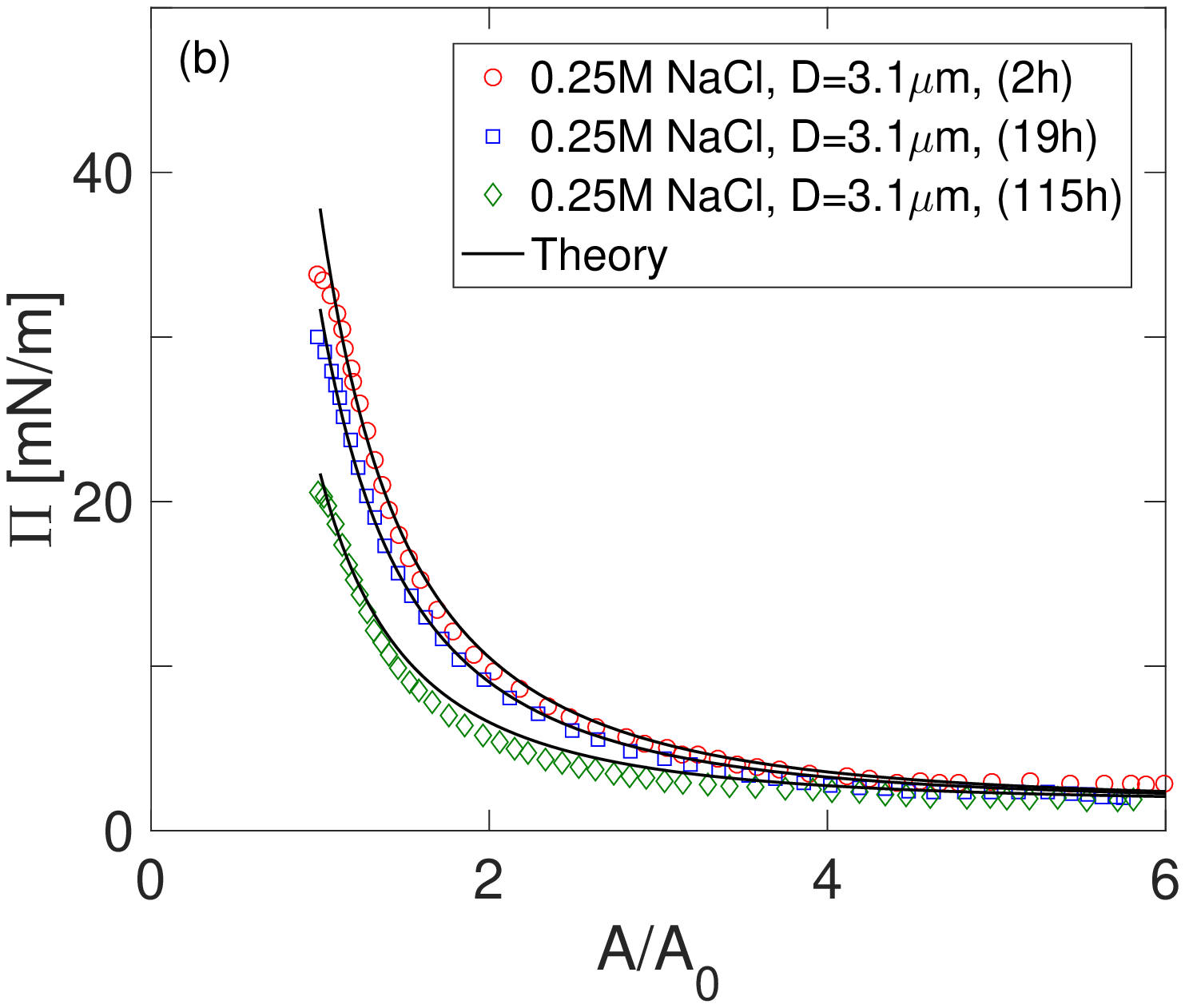} 
		\label{fig:Fig6b}
	\end{subfigure}
	\caption{\textsf{(color online) Fits of our model to experimental data. (a) Data adapted from Aveyard~\textit{et al.}~\cite{Aveyard2000}, with latex particles of radius 1.3\,$\mathrm{\mu m}$ and ionic strength of $10$\,mM spread on the octane/water interface. The fit parameter is $\sigma_{\mathrm{a}}=870$\,$\mu {\rm C}/m^2$. (b) Data adapted from Vermant~\textit{et al.}~\cite{Vermant2006}, with latex particles of radius 1.5\,$\mathrm{\mu m}$ and ionic strength of $250$\,mM spread on the decane/water interface, for different aging times that affect the surface charge. The horizontal axis, $A/A_0\sim\xi^{-2}$, is the ratio between the measured area, $A$, and its close-packing value, $A_0$. The fit parameter $\sigma_{\mathrm{a}}=720,~650$ and $530$\,$\mathrm{\mu C/m^2}$, corresponds, respectively, to aging times of $2,~19$ and $115$ hours.}}
	\label{fig:Fig6}
\end{figure}

It would be of value to compare our theoretical predictions to previous experiments. In refs~\cite{Aveyard2000} and~\cite{Vermant2006}, the surface pressure of charged polystyrene latex particles is measured at octane/water and decane/water interfaces, respectively. The fit to these experimental data shown in Figure~\ref{fig:Fig6} employs the full expression as prescribed in eqs~\ref{eq:Surface Pressure xi}, \ref{eq:Gxi_calc} and \ref{eq:g_calc}-\ref{eq:g_calc2}, and uses a single fit parameter, which is the air-exposed surface charge of a single colloid, $\sigma_{\rm a}=Q_a/[\pi (D/2)^2]$.

In general, there are two fit parameters, $\sigma_{\rm a}$ and $\sigma_{\rm w}$. However, we observe that the experiments of refs \cite{Aveyard2000,Vermant2006} used intermediate to high salt concentrations. Thus, $\kappa_{\rm D}^{-1}\leq 1$\,nm is much smaller than the colloid particle size $D\geq 1 \mu \mathrm{m}$, and $\varepsilon_{\rm a}\ll\varepsilon_{\rm w}$ for the two bounding media. It is then justified to neglect $\sigma_{\rm w}$, since the contribution from the water charges becomes much smaller (see Section VI for more details). We also find explicitly that employing $\sigma_{\rm w}$ as a second fit parameter makes no substantial difference. Nevertheless, there might exist physical scenarios where the assumption $\sigma_w=0$ is inaccurate (as is discussed below).

Figure~\ref{fig:Fig6a} presents the fit with the data of ref~\cite{Aveyard2000}. The data corresponds to an experimental setup with a $10$\,mM NaCl solution, and particles of diameter $D=2.6$\,$\mu$m. The resulting air-exposed surface charge $\sigma_{\mathrm{a}}$ is obtained as a fit parameter, $\sigma_{\rm a}=870$\,$\mathrm{\mu C/m^2}\simeq5\times10^{-3}~e/\rm nm^2$, which is a reasonable surface charge density.
We set the layer thickness $d$ to be equal to the particle diameter $D$, \textit{i.e.}, $d=D$, thus ignoring effects of colloid immersion in the aqueous phase due to wetting. The dielectric permittivities were taken to be $\varepsilon_{\mathrm{w}}=80$, $\varepsilon_{\mathrm{c}}=2.5$ and $\varepsilon_{\mathrm{oil}}=2$ for the water, polystyrene latex beads and oil (decane or octane) phases, respectively. This represents a good fit in the intermediate ionic strength regime.

In a similar fashion, Figure~\ref{fig:Fig6b} shows multiple fits to the data of ref~\cite{Vermant2006}. The surface pressure was measured for different aging times, by varying the exposure time of the monolayer in contact with $250$\,mM  NaCl solution, between 2, 19 and 115 hours. As was mentioned in ref~\cite{Vermant2006}, the number of dissociated groups on the colloid surface, corresponding to the surface charge, diminishes with time. Hence, Figure~\ref{fig:Fig6b} shows a decrease of the surface pressure with aging time.\footnote{Given that the three isotherms approach the same nonzero constant, it seems that there may be a systematic offset in the measurements. We compensate for this offset by introducing in our fit an additive constant to the surface pressure. We first fit the two-hour aging time isotherm and obtain a value of $1.7\,\mathrm{mN/m}$ for the additive offset. Then, we use this value for the two remaining isotherms.}
The colloid diameter was set to $D=3.1$\,$\mathrm{\mu m}$, and for aging times of 2, 19 and 115 hours, the fits correspond to $\sigma_{\mathrm{a}}=720,~650$ and 530\,$\mathrm{\mu C/m^2}$, respectively (charge densities of a few electrons per thousand $\rm nm^2$). As seen in Figure~\ref{fig:Fig6}, our model yields good agreement with experiments, and the fits become even better for stronger electrolytes (Figure~\ref{fig:Fig6b}), as one might expect from the DH approximation.

\begin{figure}[h!]
	\captionsetup[subfigure]{labelformat=empty}
	\begin{subfigure}{0.45\textwidth}
		\subcaption{}
		\includegraphics[width=\textwidth,draft=false, trim={0cm, 0cm, 0cm, 0cm}, clip]{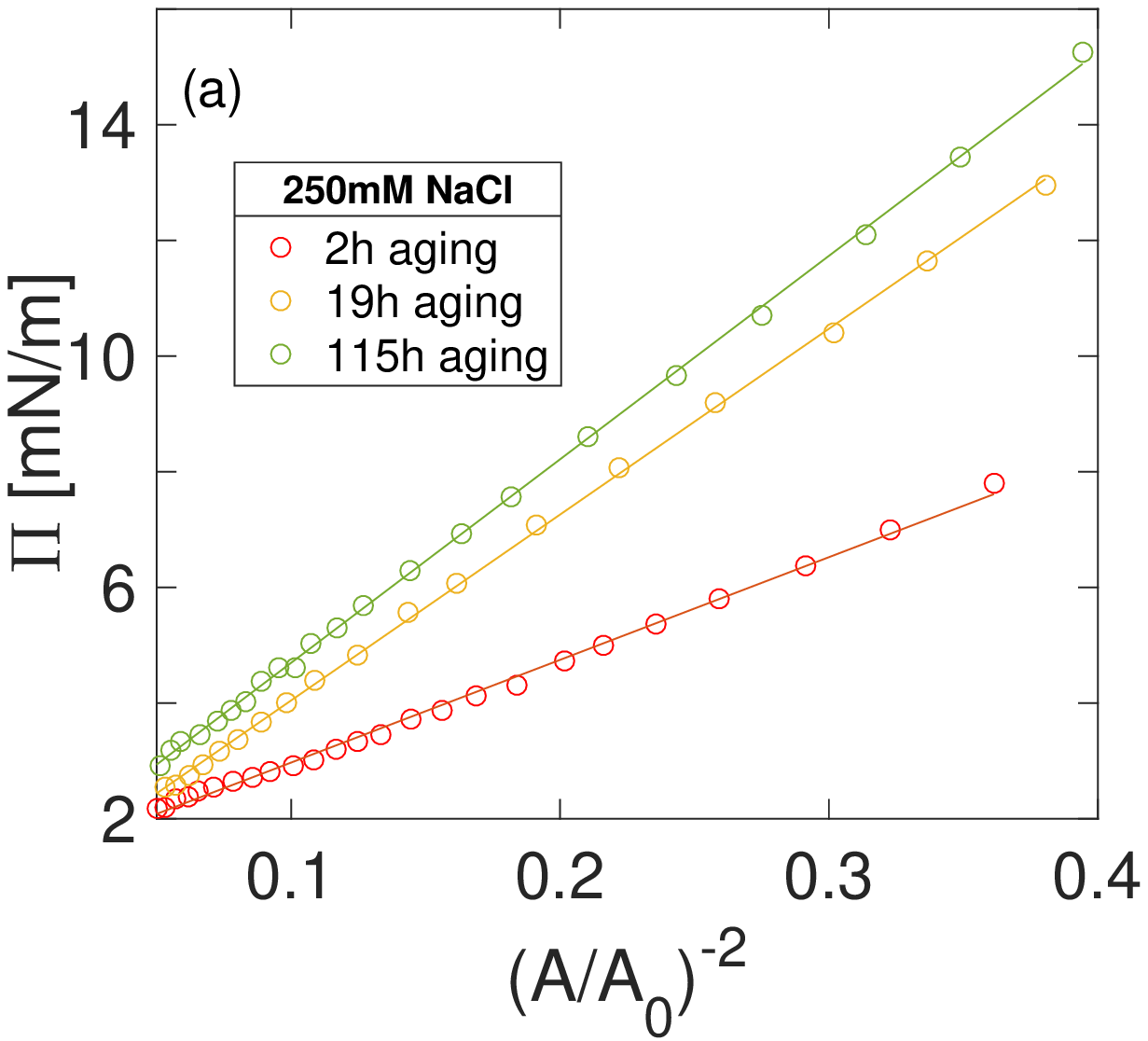}
		\label{fig:Fig7a}
	\end{subfigure}
	\begin{subfigure}{0.45\textwidth}
		\subcaption{}
		\includegraphics[width=\textwidth,draft=false,  trim={0cm, 0cm, 0cm, 0cm}, clip]{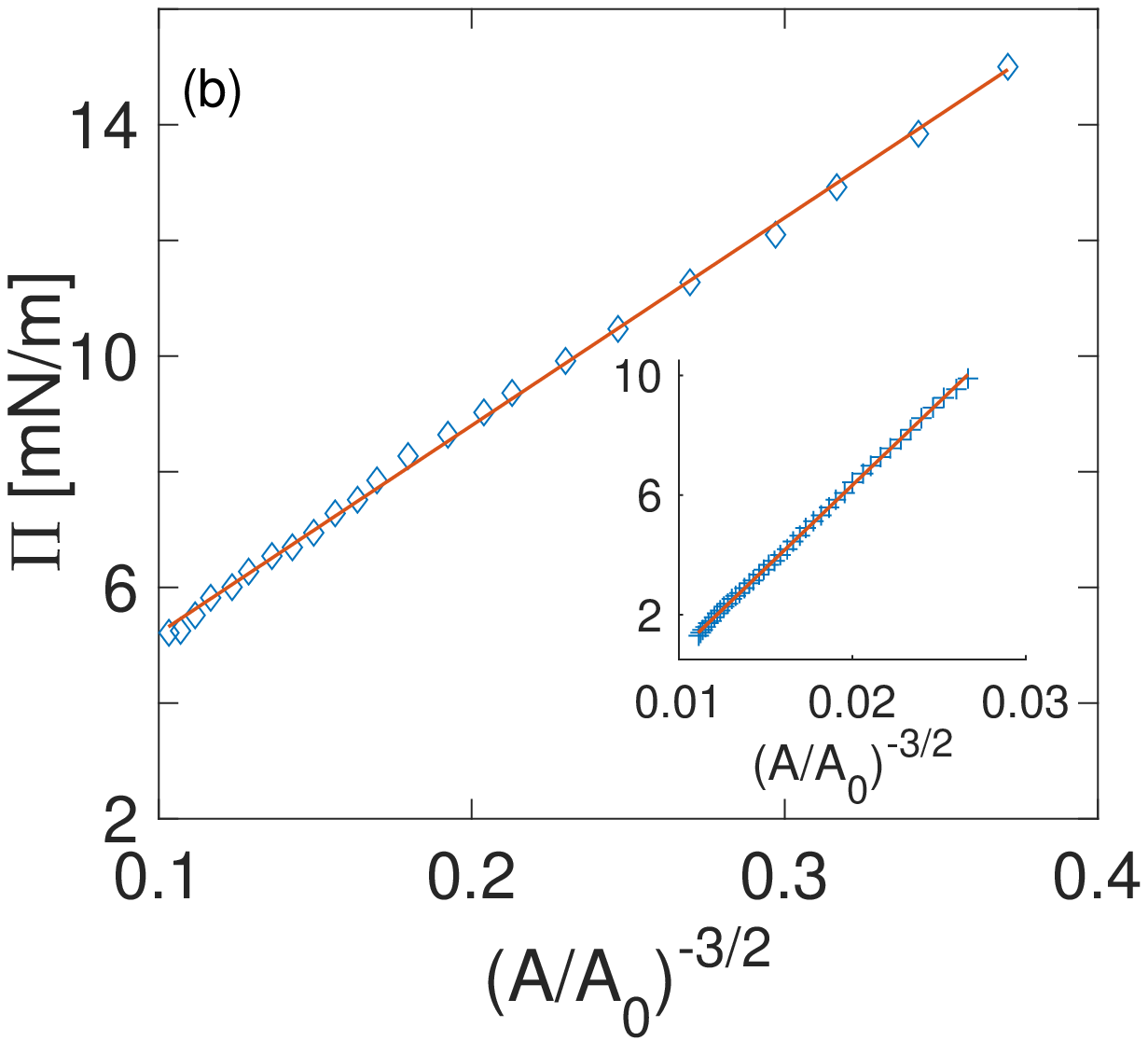} 
		\label{fig:Fig7b}
	\end{subfigure}
	\caption{\textsf{(color online) Power law fit to experimental data in different salt regimes. (a) Data adapted from Vermant {\it et al.}~\cite{Vermant2006}. The measurements were taken in the close-packing limit and in the high-salt regime, where the DH approximation holds, for aging times between 2-115 hours. For all aging times, the expected $A^{-2}$ power law agrees well with the data. (b) Data adapted from Vermant {\it et al.}~\cite{Vermant2006} (main figure) and Petkov {\it et al.} \cite{Petkov2014} (inset). Without added salt, a different power law of $(A/A_0)^{-3/2}$ agrees with the data of the same authors \cite{Vermant2006} (main figure). Measurements performed by Petkov \textit{et al} closer to the dilute limit and with no added salt also agree with the $a^{-3/2}$ power law (inset).}}
	\label{fig:Fig7}
\end{figure}

Figure \ref{fig:Fig7a} shows measurements done in the strong electrolyte case and mainly in the close packing regime~\cite{Vermant2006}. The prediction of the close-packing power law, $\Pi\sim{a}^{-2}$, follows quite well the data points. However, Figure~\ref{fig:Fig7b} shows that in the absence of salt, the measurements done in ref~\cite{Vermant2006} and by Petkov \textit{et al}~\cite{Petkov2014} exhibit a different scaling, $\Pi\sim{a}^{-3/2}$. This result is beyond the scope of our model that employs the DH approximation.

\section*{Discussion}
Scaling laws derived from the analytical results for the surface free-energy and surface pressure are obtained in two limits (see Figure~\ref{fig:Fig4}: (i)~the dilute limit ({$L \gg D$}), eq~\ref{eq:Pi_dipole}; and, (ii)~the close-packing limit ({$L \to D$}), eq~\ref{eq:Pi_cp_text}.

In the dilute limit, we have found that the surface pressure can be described in terms of dipole-dipole interactions, where the effective dipole moment $p_{\mathrm{eff}}$, eqs~\ref{eq:dipole_Qa}-\ref{eq:dipole_Qw}, arises from ionic screening in the aqueous sub-phase. The charge separation corresponding to this dipole moment, $p_{\mathrm{eff}}=p_1+p_2$, is given in terms of the colloid thickness $d$, for $p_1$ (eq~\ref{eq:dipole_Qa}) in the strong electrolyte limit, or in terms of the Debye screening length for $p_2$ (eq~\ref{eq:dipole_Qw}), in the weak limit. The sum of these two contributions, $p_1+p_2$, constitutes the effective dipole moment of each colloid. This description is valid for intermediate cases, and demonstrates an explicit dependence on the monolayer dielectric permittivity and the subphase ionic strength. We note that previous works~\cite{Aveyard2000,Oettel2008,Hurd1985} derived less general results for the dilute limit, with either $p=p_1$~\cite{Aveyard2000} or $p=p_2$ (and only for $Q_{\rm a}=0$)~\cite{Oettel2008}.

The colloidal monolayer permittivity strongly affects the surface pressure $\Pi$ through the magnitude of $p_1$, eq~\ref{eq:dipole_Qa}. The rescaled surface pressure $\Pi/\Pi_0$ of Figure~\ref{fig:Fig5} scales as $1/\varepsilon_\mathrm{c}$ in the dilute-packing limit, implying that  $\Pi\sim1/\varepsilon_{\mathrm{c}}^2$ (see eq~\ref{eq:Pi_cp_text}).  However, in the close-packing limit, the rescaled surface pressure is independent of $\varepsilon_\mathrm{c}$; hence, $\Pi\sim1/\varepsilon_{\mathrm{c}}$. The latter observation stems from the difference between the two regimes: dipole-dipole interactions vs. a uniform electric double layer.

Several comments can be made on the salt effects. First, for strong electrolytes, the surface pressure is nearly independent of ionic strength. This result is consistent with experimental findings of a weak dependence of the inter-particle force and surface pressure on salt concentration~\cite{Aveyard2000,Aveyard2002}. An exception occurs when $Q_{\rm w}$ and $Q_{\rm a}$ have opposite signs. For example, in Figure~\ref{fig:Fig3}, we plot the dependence of $\Pi$ on $\kappa_{\rm D}$ and show that it is non-monotonic and even vanishes for a specific salt concentration.

The dependence on salt is quite different for weak electrolytes. A clear divergence of the surface pressure is observed for very weak ionic strength. We remark that eq~\ref{eq:Pi_Salt} might become inaccurate in this limit, since the validity of the DH approximation breaks down for high surface charges and weak electrolytes. However, it may be more appropriate in this case to consider methods other than the DH approximation~\cite{Oettel2008,Frydel2007,Park2008}.

We would like to pay special attention to the charge on the water-side, $Q_{\rm w}$. Although it was conveniently set to zero in Figure~\ref{fig:Fig6}, we find that $Q_{\rm w}$ can, in certain cases, affect the surface pressure. Since the energetic cost of charging a surface in contact with a low dielectric material is high, it is commonly believed~\cite{Petkov2014,Bossa2016,Oettel2008} that ions from the water phase prefer to diffuse to the air-side of the layer. These ions neutralize some of the air-side charges reducing their net charge. As a result, the water-facing charge $Q_{\rm w}$  can become much higher than $Q_{\rm a}$.

In contrast to previous theoretical derivations ~\cite{Aveyard2000,Aveyard2002, Petkov2014,Bossa2016,Uppapalli2014}, which {\it a priori} neglected $Q_{\rm w}$, we can compare the contributions to the surface pressure from charges located at the top and bottom sides of the monolayer. For example, we calculate separately the surface pressure that results from charges residing only on the colloid/air interface ($\Pi_{\mathrm{ca}}$ for $Q_{\rm w}=0$), and for the opposite case, when they reside only on the colloid/water interface ($\Pi_{\mathrm{cw}}$ for $Q_{\rm a}=0$). The ratio between them, for each of the scaling limits, is given by,
\begin{equation}
\label{eq:ratio limit}
\frac{\Pi_{\mathrm{ca}}}{\Pi_{\mathrm{cw}}}= \left(\frac{Q_{\rm a}}{Q_{\rm w}} \right)^2\left(1+\frac{\varepsilon_{\rm w}}{\varepsilon_\mathrm{c}}\kappa_\mathrm{D} d\right)^{b} \,,
\end{equation}
where $b=1$ or $2$ for the close-packing and dilute regimes, respectively. Clearly, the contribution of the water-side charges cannot be neglected when the surface charge residing on the air-side is much smaller than the one on the water-side.
The corresponding $Q_{\rm w}/Q_{\rm a}$ ratio in the close-packing regime is estimated to be $10-50$ for a $1$\,mM ionic aqueous solution at room temperature and for silica particles of diameter $D=0.1-1$\,$\mu$m. Indeed, this scenario might be achieved in some physical setups.

\section*{Conclusions}

In this work we study the surface pressure of a monolayer composed of charged colloids at the air/water interface, within the linear PB theory (DH theory). The colloidal monolayer is treated as a continuum dielectric, with dielectric constant $\varepsilon_{\rm c}$ and finite thickness $d$, separating two phases: an electrolyte solution and air (or oil), with $\varepsilon_{\rm w}\gg \varepsilon_{\rm a}$. Charges of the colloidal particles facing the water-side and air-side are modeled as surface charge patches superimposed on a square lattice. As was previously suggested for similar setups~\cite{Hachisu1970, Andelman1970}, the presence of charged particles at the air/water interface results in an excess of surface free-energy, and the surface pressure can be calculated directly from the latter (see eq~\ref{eq:Surface free energy density linear}).

Although the exact solution for the surface pressure requires a numerical summation of many terms in eq~\ref{eq:Gxi_calc} (see Figure~\ref{fig:Fig4}), the scaling forms are obtained analytically, yielding for the close-packing limit ($\xi= D/L \to 1$), $\Pi\sim\xi^4\sim{a}^{-2}$ and for the dilute limit ($\xi \ll 1$), $\Pi\sim\xi^5\sim{a}^{-5/2}$. The former is consistent with the uniform surface charge density~\cite{Levental2008,Hachisu1970, Davies1951}, while the latter can be viewed as originating from dipolar interactions between discrete dipoles~\cite{Aveyard2000, Oettel2008} (see also Appendix~B).

The effective dipole moment, $p_{\rm eff}$, of the charged colloids is calculated analytically, and is found to depend on the ionic strength (Figure~\ref{fig:Fig3}), the dielectric properties of the colloidal particles (Figure~\ref{fig:Fig5}), and the amount of charges residing on the water-side ($Q_{\rm w}$) and air-side ($Q_{\rm a}$) of the colloid. We detail the physical conditions for which the contribution of the water-side charges to the surface pressure is not small, in contrast to the common belief. In addition,  the dependence on salt concentration is explored. For close-packing and dilute colloid limits, the monolayer permittivity ($\varepsilon_{\rm c}$) is shown to affect the surface pressure in different ways.

Our model agrees well with available experimental data (Figure~\ref{fig:Fig6}) using a single fit parameter, and explains the physical behavior for strong electrolytes ($\Pi\sim{a}^{-2}$ in the close-packing limit). However, some experimental results~\cite{Petkov2014,Petkov2016, Vermant2006} that exhibit longer-ranged interactions ($\Pi\sim{a}^{-3/2}$) for weak electrolytes are yet poorly understood. Our findings suggest that the latter scaling cannot be obtained within a self-consistent linear theory. This observation contradicts the theoretical model presented in \cite{Petkov2014}, where an ansatz of the linear theory was employed to describe experiments outside its range of validity (the no-salt regime). We note that a previous theoretical work \cite{Levental2008} found the proper scaling in the no-salt regime for {\it uniform} surface charge to be as strong as $\Pi\sim{a}^{-1}$, implying that the long-range scaling of the surface pressure might eventually be recovered from a fully nonlinear theory.

We hope that our study, restricted to the DH regime, will stimulate even further theoretical and experimental investigations, which will hopefully shed light on the abnormal surface pressure scaling of charged colloidal monolayers in the no-salt regime.

\vskip 0.5cm

\noindent
{\it Acknowledgements.~~} This work is supported in part by  the ISF-NSFC joint research program under Grant No. 885/15. TM acknowledges support from the Blavatnik postdoctoral fellowship programme, and DA is grateful for a Humboldt award.

\appendix
\section{Derivation of the Surface Pressure}

{We derive the solution for the electrostatic potential within the boundary value problem presented in the text. The potential $\tilde{\psi}(\mathbf{k},z)$ [with $\mathbf{k}=2\pi/(L l_{nm})$ and where $l_{nm}=\sqrt{n^2+m^2}$] in the three spatial regions ('a', 'w' and 'c' denoting air, water, and colloid, respectively) is obtained from eqs~\ref{eq:Laplace Equation1}, ~\ref{eq:DH Equation}, and \ref{eq:Periodic Potential Fourier}:}

\begin{eqnarray}
\label{eq:Vector Psi1}
& &{\tilde{\psi}^{(\mathrm{a})}(\mathbf{k},z)=\tilde{\psi}_{\mathrm{s}}^{(\mathrm{a})}(\mathbf{k})\exp\left(-\Lambda_{nm} z\right) \, ,}
\nonumber\\
& &{\tilde{\psi}^{(\mathrm{w})}(\mathbf{k},z)=\tilde{\psi}_{\mathrm{s}}^{(\mathrm{w})}(\mathbf{k},z)}
\nonumber \\
& & {\qquad\qquad\,\,\,\times\exp\left[ {\left(\Lambda_{nm}^2+\kappa_\mathrm{D}^2\right)}^{1/2} (z+d)\right] \, ,} \nonumber\\
& &{\tilde{\psi}^{(\mathrm{c})}(\mathbf{k},z)=\tilde{\psi}_{\mathrm{s}}^{(\mathrm{a})}(\mathbf{k})\frac{\sinh[\Lambda_{nm} (z+d)]}{\sinh(\Lambda_{nm} d)}}
\nonumber \\
& &{\qquad\qquad\,\,\,-\tilde{\psi}_{\mathrm{s}}^{(\mathrm{w})}(\mathbf{k})\frac{\sinh(\Lambda_{nm} z)}{\sinh(\Lambda_{nm} d)}\, ,}
\end{eqnarray}
{with $\Lambda_{nm}\equiv 2\pi \xi l_{nm}/D=2\pi l_{nm}/L$. Employing the boundary conditions, eq~\ref{eq:Vector_BC1_eq}, we obtain for the four capacitance matrix elements $\mathrm{C}^{-1}_{ij}$ (eq~\ref{eq:Vector Linear Relation between surface potential and surface charge}):}
\begin{align}
\label{eq:Capacitance Matrix}
\begin{split}
{\mathrm{C^{-1}_{11}}}&{= \frac{1}{c_{nm}}\Bigg[1+\frac{\varepsilon_{\rm w}}{\varepsilon_\mathrm{c}} \sqrt{1+\left(\kappa_\mathrm{D}/\Lambda_{nm}\right)^2} \tanh\left(\Lambda_{nm} d\right)\Bigg]\, ,} \\
{\mathrm{C^{-1}_{22}}}&{=\frac{1+(\varepsilon_{\rm a}/\varepsilon_\mathrm{c}) \tanh\left(\Lambda_{nm} d\right)}{c_{nm}} \, ,} \\
{\mathrm{C^{-1}_{12}}}&{=\mathrm{C^{-1}_{21}}=\frac{1}{c_{nm}}\frac{1}{\cosh\left(\Lambda_{nm} d\right)} \, ,}
\end{split}
\end{align}
{with $c_{nm}$ above given by
\begin{equation}
\label{eq:cnm}
\begin{split}
{c_{nm}}&{=\varepsilon_0\varepsilon_{\rm w} \Lambda_{nm}\Bigg[\sqrt{1+(\kappa_\mathrm{D}/\Lambda_{nm})^2}+\frac{\varepsilon_{\rm a}}{\varepsilon_\mathrm{w}}}\\&
{+\left(\frac{\varepsilon_{\rm a}}{\varepsilon_\mathrm{c}}\sqrt{1+(\kappa_\mathrm{D} /\Lambda_{nm})^2}+\frac{\varepsilon_{\rm c}}{\varepsilon_{\rm w}}\right)\,\tanh\left(\Lambda_{nm}d\right)\Bigg]}.\\
\end{split}
\end{equation}}

We now turn to develop the mathematical framework required for the derivation of the surface free energy and pressure. Using the notation $0\le \xi\equiv D/L\le 1$, eq~\ref{eq:Vector Surface Pressure Explicit 1} is written as

\begin{equation}
\label{eq:Gxi}
{\braket{\Delta f_{\rm el}}\equiv \frac{2\xi^4}{D^4}\sum_{n,m=0}^{\infty}\theta_{nm}(\tilde{\Sigma}\mathrm{C^{-1}}\tilde{\Sigma})_{nm} \, ,}
\end{equation}
{where $\theta_{nm}=(2-\delta_{n0})(2-\delta_{m0})/4 \,$ depends on the Kronecker delta function, $\delta_{nm}$, and where we made use of the square lattice symmetry of our setup. Finally, from eqs~\ref{eq:Surface Pressure Non-Uniform} and \ref{eq:Gxi}, the surface pressure is given by}
\begin{equation}
\label{eq:Surface Pressure xi}
{\Pi= \frac{1}{2}\xi\frac{d \braket{\Delta f_{\mathrm{el}}}}{d\xi} -\braket{\Delta f_{\mathrm{el}}} \, .}
\end{equation}

{The Fourier transform of $s(r)$ of eq~\ref{eq:Gauss} has a Gaussian form,}
\begin{equation}
\label{eq:Gauss_FT}
{\tilde{s}(n\xi,m\xi)=Q\exp{\left[-\frac{\pi^2}{2}\xi^2l_{nm}^2\right]} \, ,}
\end{equation}
{Using this expression, the explicit solution for the electrostatic potential and the capacitance matrix (eqs \ref{eq:Vector Psi1} and \ref{eq:Capacitance Matrix}), we write}
\begin{equation}
\label{eq:Gxi_calc}
\begin{split}
{\braket{\Delta f_{\mathrm{el}}}}&{=\frac{g(0)}{2}\xi^4+2\xi^4\sum_{n=1}^{\infty} g(\xi l_{n0})}\\&{+2\xi^4\sum_{n,m=1}^{\infty} g(\xi l_{nm}) \, .}
\end{split}
\end{equation}
{where $g(\rho)$ is a radially symmetric function ($\rho=\xi l_{nm}$), found from eqs \ref{eq:Vector Psi1}-\ref{eq:cnm},}
\begin{equation}
\label{eq:g_calc}
\begin{split}
{g(\rho)}&{=\frac{1}{2 \pi D^3\varepsilon_0\varepsilon_{\rm w}} \frac{1}{\rho h(\rho)}
	\rm e^{-\pi^2\rho^2}}\\ &{\times\Big(Q_{\rm a}^2\Big[1+\frac{\varepsilon_{\rm w}}{\varepsilon_\mathrm{c}}\sqrt{1+(\kappa_\mathrm{D} D/2\pi\rho)^2} \tanh\left(2\pi \rho d/D\right)\Big]}
\\&{+Q_{\rm w}^2\Big[1+\frac{\varepsilon_{\rm a}}{\varepsilon_\mathrm{c}}\tanh\left(2\pi \rho d/D\right)\Big]}
\\&{+\frac{2Q_{\rm a}Q_{\rm w}}{\cosh\left(2\pi \rho d/D\right)}\Big) \, ,}
\end{split}
\end{equation}
{with $h(\rho)$ above defined as
\begin{equation}
\label{eq:g_calc2}
\begin{split}
{h(\rho)}&{=\sqrt{1+(\kappa_\mathrm{D} D/2\pi\rho)^2}+\frac{\varepsilon_{\rm a}}{\varepsilon_\mathrm{w}}}
\\&{+\left(\frac{\varepsilon_{\rm a}}{\varepsilon_\mathrm{c}}\sqrt{1+(\kappa_\mathrm{D} D/2\pi\rho)^2}+\frac{\varepsilon_{\rm c}}{\varepsilon_\mathrm{w}}\right)} \\ &{\times\tanh\left(2\pi \rho d/D\right) \, .}
\end{split}
\end{equation}
For the analytic derivation of the surface pressure regimes given below, it is sufficient to use $g(\rho)$ and $g'(\rho)$ evaluated at the origin, $\rho=0$,
\begin{equation}
\label{eq:g_0}
{g(0)=\frac{Q_{\rm a}^2d}{\varepsilon_\mathrm{c}\varepsilon_0 D^4}+\frac{(Q_{\rm a}+Q_{\rm w})^2}{\varepsilon_0\varepsilon_{\rm w}\kappa_\mathrm{D} D^4} \, ,}
\end{equation}
and
\begin{equation}
\label{eq:g'0}
g'(0)=-\frac{\varepsilon_{\rm a}}{\varepsilon_\mathrm{c}}\frac{2\pi d^2}{\varepsilon_\mathrm{c}\varepsilon_0 D^5} \left[Q_{\rm a}+\frac{\varepsilon_\mathrm{c}}{\varepsilon_{\rm w}}\frac{Q_{\rm a}+Q_{\rm w}}{\kappa_\mathrm{D} d} \right] ^2 \, .
\end{equation}
}
{In the close-packing limit, $\Pi$ can be derived analytically by investigating $\braket{\Delta f_{\mathrm{el}}}$ and its derivative in the $\xi=D/L \to  1$ limit. The dominant contribution to eq~\ref{eq:Gxi_calc} originates from the first term, since a simple substitution of $\xi\to1$ in eq~\ref{eq:Gxi_calc} implies that the contributions from the two remaining series are exponentially small, approximately of order $\exp{(-\pi^2)}\sim 10^{-5}$, and can be safely neglected. Then, from eqs~\ref{eq:Gxi_calc} and \ref{eq:g_0} one can derive}
\begin{equation}
\label{eq:F_cp}
{\braket{\Delta f_{\mathrm{el}}}\simeq \frac{1}{2}\left[\frac{Q_{\rm a}^2d}{\varepsilon_\mathrm{c}\varepsilon_0}+
\frac{(Q_{\rm a}+Q_{\rm w})^2}{\varepsilon_0\varepsilon_{\rm w}\kappa_\mathrm{D}}\right]\frac{\xi^4}{D^4} \,.}
\end{equation}
{Eq~\ref{eq:Pi_cp_text} of the text is obtained from
eqs~\ref{eq:Surface Pressure xi} and \ref{eq:F_cp},
and by the substitution $(\xi/D)^4=L^{-4}= a^{-2}$.}

{For the dilute limit, $\xi=D/L \ll 1$, we remark that $\braket{\Delta f_{\mathrm{el}}}$ has the form of a Riemann sum \cite{Abramowitz}. Following this observation, it can be evaluated in the limit of small $\xi$. For convenience we express it as (see eq~\ref{eq:Gxi_calc})}
	\begin{equation}
	\label{eq:H_xi}
	\begin{split}
	{\frac{\braket{\Delta f_{\mathrm{el}}}}{2\xi^2}}&{=\frac{g(0,0)}{4}\xi^2+\xi \sum_{n=1}^{\infty}g(n\xi,0)\xi}
	\\&{+\sum_{n,m=1}^{\infty}g(n\xi,m\xi)\xi^2 \, .}
	\end{split}
	\end{equation}
	{Here, $g(x,y)$ is a general well-behaved function of two variables. For the one-dimensional sum, we employ the Euler-Maclaurin formula~\cite{Abramowitz},}
	\begin{equation}
	\label{eq:EM_1d}
	\begin{multlined}
	{\sum_{n=1}^{\infty}g(n\xi,0)\xi=\sum_{k=0}^{\infty}\frac{B_k}{k!}
		\xi^k \int_0^\infty\frac{\partial^{k}g}{\partial x^{k}}\ \Big|_{y=0}\D x \,,}
	\end{multlined}
	\end{equation}
	{where $B_k$ are the Bernoulli numbers, with the first five given by
		$$B_0=1, B_1=\frac{1}{2}, B_2=\frac{1}{6}, B_3=0, B_4 =-\frac{1}{30} \, .$$}
	{If $\lim_{x\to \infty}\partial^kg/\partial x^k=0$ for all $k$,
		eq~\ref{eq:EM_1d} implies that the expansion of $\sum_{n=1}^{\infty}g(n\xi,0)\xi$ in powers of $\xi$ is}
	\begin{equation}
	\label{eq:EM_1d_exp}
	\begin{split}
	{\sum_{n=1}^{\infty}g(n\xi,0)\xi}&{=\int_0^{\infty}g(x,0)\,\D x\,-\frac{g(0,0)}{2}\xi}
	\\&{-~\frac{g'(0,0)}{12}\xi^2+...}
	\end{split}
	\end{equation}

	{For the double sum, the generalization of the Euler-Maclaurin formula is given by}
	\begin{equation}
	\label{eq:EM_2d}
	\begin{split}
	{\sum_{n,m=1}^{\infty}g(n\xi,m\xi)\xi^2}
	&{=\sum_{j,k=0}^{\infty} \frac{B_j}{j!}\frac{B_k}{k!} \xi^{j+k}}\\ &{\times
		\int_0^\infty\int_0^\infty \frac{\partial^{j}\partial^{k}g}{\partial x^{j}\partial y^{k}}\,\D x\D y \, ,}
	\end{split}
	\end{equation}
	{If $g(x,y)$ has radial symmetry, it can be written as $g(x,y)=g(\rho)$ where $\rho=\sqrt{x^2+y^2}$, coinciding with the expression of $g$ in eq~\ref{eq:g_calc}, where $\rho=\xi l_{nm}$. We can then calculate the double integral in polar coordinates ($\rho,\theta$), while recalling that $g$ and all of its $\rho$-derivatives should vanish at infinity. These assumptions lead to the following expansion for the double sum,}

	\begin{equation}
	\label{eq:EM_2d_exp}
	\begin{split}
	{\sum_{n,m=1}^{\infty}g(n\xi,m\xi)\xi^2}
	&{=\int_0^{\pi/2}\D\theta\int_0^{\infty}g(\rho)\rho \,\D \rho}\\
	&{\,-~\xi\int_0^{\infty}g(\rho)\,\D \rho\,+\,\frac{g(0)}{4}\xi^2}\\
	&{\,+\,\frac{g'(0)}{36}\xi^3+...}
	\end{split}
	\end{equation}

	{Finally, substituting eqs~\ref{eq:EM_1d_exp} and \ref{eq:EM_2d_exp} into eq~\ref{eq:H_xi}, we obtain the expansion of $\braket{\Delta f_{\mathrm{el}}}$ for vanishing $\xi$,
\begin{equation}
\label{eq:G_dilute}
\braket{\Delta f_{\mathrm{el}}}\simeq 2\xi^2\int_0^{\pi/2}\D\theta \int_0^{\infty} g(\rho)\,\rho\,\D \rho \,-\,\, \frac{g'(0)}{9}\xi^5 \, .
\end{equation}
Using eqs~\ref{eq:Surface Pressure xi}, \ref{eq:g'0} and \ref{eq:G_dilute}, the value of $\Pi$ in the dilute limit is obtained as
\begin{equation}
\label{eq:Pi_lp}
{\Pi(\xi)= \frac{\pi}{3}\frac{\varepsilon_{\rm a}}{\varepsilon_\mathrm{c}}\frac{d^2}{\varepsilon_\mathrm{c}\varepsilon_0} \left[Q_{\rm a}+\frac{\varepsilon_\mathrm{c}}{\varepsilon_{\rm w}}\frac{Q_{\rm a}+Q_{\rm w}}{\kappa_\mathrm{D}d} \right] ^2 \left(\frac{\xi}{D}\right)^5 \, .}
\end{equation}
Eq~\ref{eq:Pi_dipole} of the text is recovered by substituting $(D/\xi)^2=a$.}

\section{Surface pressure and dipole interactions for $\xi\ll1$}
{We present another way to obtain eqs~\ref{eq:Pi_dipole}-\ref{eq:dipole}. We start by considering the interaction energy between two parallel point-like dipoles of magnitude $p$ at a large separation $L$ that is perpendicular to the dipole direction. The dipoles are placed in the upper half-space having permittivity $\varepsilon_{\rm a}$. Assuming that there is no contribution from the lower half-space (with permittivity $\varepsilon_{\rm w}\gg \varepsilon_{\rm a}$), the dipole-dipole interaction energy is~\cite{Oettel2008}
\begin{equation}
\label{eq:dipole_int}
u_{\mathrm{int}}=\frac{p^2}{8\pi\varepsilon_{\rm a}\varepsilon_0L^3} \, ,
\end{equation}
and is equal to one half of the familiar expression for interacting parallel dipoles. Summing over all pairs of parallel dipoles placed on a square lattice embedded in 3D space, the lattice cohesive energy, $U_{\mathrm{c}}$, is given by
\begin{eqnarray}
\label{eq:dipole_energy}
\begin{split}
&U_{\mathrm{c}} = \frac{N}{2} \sum_{\boldsymbol{\rho}_{n m}\neq 0}u_{\mathrm{int}}(\boldsymbol{\rho}_{n m}) \\&\simeq
0.18 N^{5/2} \frac{p^2}{\varepsilon_{\rm a}\varepsilon_0 A^{3/2}} \, ,
\end{split}
\end{eqnarray}
where $\boldsymbol{\rho}_{n m}=L(n,m)$ is an in-plane lattice vector, $N$ the number of particles (and lattice sites), and $A=NL^2$ the total surface area. Similar to the derivation of eq~\ref{eq:Surface Free Energy DLVO}, we neglect the entropy of the large colloidal particles. Moreover, the entropy of the mobile electrolyte ions approaches the bulk-solution value, which is independent of the area, yielding $U_{\mathrm{c}}=F_{\mathrm{el}}+\rm const.$}

{In eq~\ref{eq:dipole_energy}, we performed a summation over all lattice sites,
$$ \sum_{(m,n)\neq(0,0)} \frac{1}{(m^2+n^2)^{3/2}}\simeq 9.03\, .$$
The surface pressure is then recovered by taking $\Pi= -\mathrm{d}F_{\mathrm{el}}/\mathrm{d}A\simeq -\mathrm{d}U_{\rm c}/\mathrm{d}A$,
\begin{equation}
\label{eq:dipole_pi}
\begin{multlined}
\Pi\simeq 0.269 \frac{p^2}{\varepsilon_{\rm a}\varepsilon_0 a^{5/2}} \, .
\end{multlined}
\end{equation}
Comparing eq~\ref{eq:dipole_pi} with eq~\ref{eq:Pi_dipole}, we find the $p$ value as calculated above is $p=0.99p_{\rm eff}$ of the $p_{\rm eff}$ in eq~\ref{eq:Pi_dipole}. Hence, in the dilute limit, eq~\ref{eq:dipole} can be regarded as the effective dipole moment of the colloidal particle.}


\end{document}